\documentclass[prx,aps,amsmath,amssymb,showpacs,twocolumn]{revtex4-2}
\usepackage{graphicx,amsmath,amssymb,color}
\usepackage[utf8]{inputenc}
\usepackage{nicefrac}
\usepackage{multirow, makecell, ragged2e}
\usepackage[titletoc,title]{appendix}
\usepackage{physics}
\usepackage[colorlinks,bookmarks=true,citecolor=blue,linkcolor=red,urlcolor=blue]{hyperref}

\newcommand{\be}{\begin{equation}}
\newcommand{\ee}{\end{equation}}

\newcommand{\ba}{\begin{eqnarray}}
\newcommand{\ea}{\end{eqnarray}}

\renewcommand{\vec}[1]{\mbox{\boldmath$#1$}}

\def\beq{\begin{eqnarray}}
\def\eeq{\end{eqnarray}}
\newcommand{\sh}{\mathcal{S}}

\newcommand{\elliptic}[5][\scriptstyle]{\vartheta\left[\begin{array}{c}{{#1 #2}}\\{#1 #3}\end{array}\right]\left(#4\middle|#5\right)}

\begin{document}

\title{ 
Composite-fermion pairing at half-filled and quarter-filled lowest Landau level}
\author{Anirban Sharma,$^1$ Ajit C. Balram,$^{2,3}$ and J. K. Jain$^1$}
\affiliation{$^1$Department of Physics, 104 Davey Lab, Pennsylvania State University, University Park, Pennsylvania 16802,USA}
\affiliation{$^2$Institute of Mathematical Sciences, CIT Campus, Chennai, 600113, India}
\affiliation{$^3$Homi Bhabha National Institute, Training School Complex, Anushaktinagar, Mumbai 400094, India}
\date{\today}

\begin{abstract} 
The Halperin-Lee-Read Fermi sea of composite fermions at half-filled lowest Landau level is the realization of a fascinating metallic phase that is a strongly correlated ``non-Fermi liquid" from the electrons' perspective. Remarkably, experiments have found that as the width of the quantum well is increased, this state makes a transition into a fractional quantum Hall state, the origin of which has remained an important puzzle since its discovery more than three decades ago.  We perform detailed and accurate quantitative calculations using a systematic variational framework for the pairing of composite fermions that closely mimics the Bardeen-Cooper-Schrieffer theory of superconductivity.  Our calculations show that (i) as the quantum-well width is increased, the single-component composite-fermion Fermi sea occupying the lowest symmetric subband of the quantum well undergoes an instability into a single-component $p$-wave paired state of composite fermions; (ii) the theoretical phase diagram in the quantum-well width - electron density plane is in excellent agreement with experiments; (iii) a sufficient amount of asymmetry in the charge distribution of the quantum well destroys the fractional quantum Hall effect, as observed experimentally; and (iv) the two-component 331 state is energetically less favorable than the single component paired state. Evidence for fractional quantum Hall effect has been seen in wide quantum wells also at quarter-filled lowest Landau level; here our calculations  indicate an $f$-wave paired state of composite fermions. We further investigate bosons in the lowest Landau level at filling factor equal to one and show that a $p$-wave pairing instability of composite fermions, which are bosons carrying a single flux quantum, occurs  for the short range as well as the Coulomb interaction, in agreement with exact diagonalization studies. The general consistency of the composite-fermion Bardeen-Cooper-Schrieffer approach with experiments lends support to the notion of composite-fermion pairing as the mechanism of fractional quantum Hall effects at even-denominator filling factors. Various experimental implications are mentioned.
\end{abstract}
\maketitle

\section{Introduction}

The observation of a fractionally quantized Hall plateau at $R_H=h/\nu e^2$ indicates the formation of an incompressible state at filling fraction $\nu$~\cite{Laughlin83}. Beginning with $\nu=1/3$~\cite{Tsui82}, a large array of fractions have been observed~\cite{DasSarma07,Halperin20}. Most of the observed fractions have the form $\nu=n/(2pn\pm 1)$, $n$ and $p$ integers, which are understood as the integer quantum Hall effect of composite fermions (CFs), namely  electrons bound to an even number ($2p$) of quantized vortices~\cite{Jain89,Jain20}. These fractions terminate into compressible states at even-denominator fractions such as $\nu=1/2$, which are realizations of the Fermi seas of CFs~\cite{Halperin93,Halperin20,Halperin20b,Shayegan20}.
The first even-denominator fractional quantum Hall effect (FQHE) was observed at $\nu=5/2$ in GaAs quantum wells~\cite{Willett87}. Moore and Read (MR) proposed a Pfaffian (Pf) state~\cite{Moore91}, which was subsequently interpreted as representing a $p$-wave pairing of CFs and associated with the $\nu=5/2$ FQHE~\cite{Greiter91, Greiter92a, Read00, Morf98, Scarola02b, Balram20b}. This state is akin to topological superconductivity of CFs and is therefore believed to host quasiparticles obeying non-Abelian statistics~\cite{Moore91,Read00}. More recently, Balram, Barkeshli and Rudner~\cite{Balram18} have shown that the 5/2 state can also be successfully modeled in terms of the so-called ``$\bar{2}\bar{2}111$" parton wave function, which belongs to the class of wave functions introduced in Ref.~\cite{Jain89b} and shown in Ref.~\cite{Wen91} to host non-Abelian excitations.  

M\"oller and Simon~\cite{Moller08} and Sharma {\it et al.}~\cite{Sharma21} have treated the CF pairing in the $5/2$ state in an approach that closely mimics the Bardeen-Cooper-Schrieffer (BCS) theory of superconductivity and shown that the CF Fermi sea (CFFS) is unstable to pairing of CFs in the $p$-wave channel. As with the BCS theory, this approach can be used to provide a unified treatment of pairing instabilities in different relative angular momentum channels and to make predictions regarding the optimal pairing channel. Also, because it contains the CFFS as a limiting case, it can in principle be applied to situations where a transition occurs, as a function of some parameter, from the compressible CFFS state into a paired FQHE state. 

While a FQHE has been observed at $\nu=5/2$ in the second Landau level (LL), the states at $\nu=1/2$ and $\nu=1/4$ in narrow quantum wells (QWs) are well established to be compressible Fermi seas of CFs carrying two and four vortices, respectively~\cite{Halperin93,Halperin20b,Shayegan20,Willett93,Kang93,Goldman94,Smet96,Smet98a,Smet99,Park98b, Balram15c, Kamburov14,Hossain19}, as expected for weakly interacting CFs. Unexpectedly, Suen {\em et al.} observed FQHE at $\nu=1/2$ in wide QWs in 1992~\cite{Suen92,Suen94}, followed by systematic studies demonstrating that a transition from the CFFS to a FQHE state occurs as the width of the QW or the electron density is increased~\cite{Luhman08,Shabani13}. A similar behavior was observed at $\nu=1/4$ by Shabani and collaborators~\cite{Shabani09a,Shabani09b,Shabani13}.  One may ask why electrons at $\nu=1/2$ and $\nu=1/4$ in the lowest LL (LLL) behave differently from $\nu=5/2$ in the second LL in narrow QWs, and why a FQHE state appears at these LLL filling factors in wide QWs? The fate of an even-denominator state eventually depends on the interaction between CFs, which is a remnant of the electron-electron (e-e) interaction after a large part of it is exhausted in forming the CFs. The interaction between CFs is a complex function of the parent e-e interaction, and it can sometimes be estimated numerically within an approximate scheme~\cite{Sitko96, Lee01, Lee02, Wojs04, Balram16c, Balram17b}. The picture that has emerged from numerical studies and by comparing the CF theory to experiments is that the model of noninteracting CFs is qualitatively valid when the e-e interaction is strongly repulsive at short distance, as is the case in the LLL for narrow QWs. When the e-e interaction is made less repulsive at short distance, the interaction between CFs can become attractive, which may lead to a pairing instability. There are several ways of making the e-e interaction less repulsive at short distance: by going to higher LLs, by making QWs wider, and by enhancing LL mixing. Of course, as the e-e interaction strength is reduced at short distance, eventually CFs themselves become unstable, and some other state, such as a charge density wave of electrons, takes over. A paired-CF state is obtained in a sweet spot where the e-e interaction is sufficiently strongly repulsive as to produce CFs, but not so strongly repulsive as to produce a CFFS. It is thus a delicate quantitative question whether CF pairing occurs, and if so, in what channel.

\begin{table*}[t]
\begin{center}
 \begin{tabular}{|l |l |c |c|} 
 \hline
pairing symmetry  &    candidate states  &  shift $\mathcal{S}=l+s$ & central charge $c=1+l/2$ \\
\hline\hline 
$l=3$ ($f$-wave)     &    MR-$f$, CF-BCS-$f$, 221$^{s-1}$  & $3+s$  & $5/2$~\cite{Wen91} \\
$l=1$ ($p$-wave)     &    MR-Pf, CF-BCS-$p$, anti-($\bar{2}\bar{2}111)1^{s-2}$  & $1+s$ & $3/2$~\cite{Moore91} \\
$l=-1$     &    PHS-Pf, CF-BCS$_{l=-1}$  & $-1+s$ & $1/2$~\cite{Son15} \\
$l=-3$    &    anti-Pf, CF-BCS$_{l=-3}$, $\bar{2}\bar{2}1^{s+1}$  & $-3+s$ & $-1/2$~\cite{Balram19} \\
\hline
  \end{tabular}
\end{center}
\caption{\label{table-exp} This table lists the candidate states at $\nu=1/s$ for various pairing channels, where $s$ is an even integer for fermions and an odd integer for bosons. The term ``anti" refers to hole conjugation. The values of the shift $\mathcal{S}$ and the chiral central charge $c$ are also given.}
\end{table*}

We will address the issue from the CF-BCS perspective. In this approach, we proceed by constructing a BCS wave function for CFs at $\nu=1/2$ and $\nu=1/4$ in the periodic torus geometry~\cite{Sharma21}. For this purpose, we composite fermionize the BCS wave function for electrons by attaching even number of quantized vortices to electrons and then projecting it into the LLL~\cite{Jain89}. The CF-BCS wave function has two variational parameters, which are analogous to the gap function and the Debye cutoff in the standard BCS theory. While the idea is in principle straightforward, its implementation is nontrivial, because several technical hurdles must be overcome, and in particular, we must modify the Jain-Kamilla (JK) projection scheme \cite{Jain97,Jain97b}, because the standard JK projection in the torus geometry takes us out of the original Hilbert space by producing unphysical wave functions that do not satisfy the stipulated periodic boundary conditions (PBCs).

The CF BCS approach offers several advantages. First of all, this method is not tied to a specific wave function, but casts a wider net where one can search for the lowest energy wave function of the BCS form by varying two parameters. (The MR and parton wave functions do not contain any variational parameters.) Second, the CF-BCS wave function reduces to the CFFS in one limit, which will be the lowest energy solution when a pairing instability is absent. This method thus can tell if the compressible CFFS is more likely to occur than an incompressible paired state. It is expected to be the most reliable when the CFFS is a good starting point, which is the case at $\nu=1/2$ and $\nu=1/4$ in the LLL. Finally, this method also enables a study of the competition between different pairing channels.  A mention of the limitations of the method is also in order. It is important to work with CFFS configurations that are nearly circular, which limits our study to only a few particle numbers (such as $N=12$, $32$, and $60$ electrons for a square torus). We will neglect LL mixing, which may also induce pairing~\cite{Wang22,Zhao23}. In the end, we note that we will also not consider in this work certain other competing states, such as stripes or Wigner crystal.

In this article, we apply the CF-BCS method to several situations where even-denominator FQHE has been observed. To set our convention, we work with a gap function $\Delta^{(l)}_{\vec{k}}\sim e^{-il\theta}$, where $\theta$ is the angular coordinate of the wave vector $\vec{k}$. We consider pairing channels with relative angular momentum $l=3$ (denoted below as $f$-wave), $l=1$ ($p$-wave), $l=-1$ and $l=-3$. The state with the pairing channel $l=3$ lies in the same universality class as the ``$221$" parton state \cite{Balram18}; $l=1$ belongs in the same phase as the MR-Pf or the anti-$\bar{2}\bar{2}111$-parton states; $l=-3$ paired state lies in the same phase as the anti-Pfaffian~\cite{Levin07, Lee07} or the $\bar{2}\bar{2}111$-parton states \cite{Balram18}; and $l=-1$ paired state is topologically equivalent to the so-called particle-hole symmetric Pf (PHS-Pf) phase~\cite{Son16}. (Here, ``anti" denotes hole conjugate state.) These states and some of their topological quantum numbers are listed in Table~\ref{table-exp}. We provide here a brief summary of our results, including, for completeness, the conclusions from earlier work~\cite{Sharma21,Sharma23}.  Further details are given in subsequent sections.

\underline{$\nu=1/2$ and $5/2$ in narrow QWs:} It was shown by Sharma {\it et al.}~\cite{Sharma21} that, for a two-dimensional (2D) system, no pairing instability is seen at $\nu=1/2$, whereas a $p$-wave instability occurs at $\nu=5/2$. This is consistent with experiments in the sense that FQHE has been seen at $\nu=5/2$ but not at $\nu=1/2$. These results are applicable, within approximations, to even denominator FQHE observed in bilayer graphene \cite{Ki14,Kim15,Zibrov17,Li17, Huang22, Balram21b, Assouline23, Hu23}, which occurs in Landau bands that are analogous to the second LL of GaAs QWs. 

\underline{Even denominators in monolayer graphene:}  FQHE at even-denominator fractions has been observed by Kim {\it et al.} in the $\mathcal{N}=3$ LL of monolayer graphene~\cite{Kim19}. The CF-BCS formulation finds an $f$-wave pairing instability for the interaction appropriate for this LL~\cite{Sharma23}. This result is consistent with the calculations in Ref.~ \cite{Kim19} which had suggested that the 221 parton state~\cite{Jain89b}, which also represents an $f$-wave pairing~\cite{Wen91, Balram18}, was the most plausible incompressible state among various states considered.

\underline{$\nu=1/2$ in wide QWs:} FQHE has been seen at $\nu=1/2$ in wide QWs~\cite{Suen92,Suen94,Luhman08,Shabani13}. Whether the observed FQHE state is a one or a two component state has been a matter of debate for three decades \cite{Greiter92,He93,Suen94b,Peterson10,Liu14d,Thiebaut15,Mueed15,Mueed16}, where the two components here would be the lowest two subbands whose separation becomes small with increasing width. We begin by assuming a single-component origin, determine the effective e-e interaction as a function of the QW width and the electron density and find that the CFFS is unstable to $p$-wave pairing at large QW widths and / or large densities. Our phase diagram is in excellent agreement with the experimentally determined phase diagram in the QW width - density plane. We also find, in agreement with experiments, that the FQHE in wide QWs is destroyed when the charge distribution in the QW is made sufficiently asymmetric. Our calculations also indicate that the two-component Halperin 331 state is energetically less favored. These results demonstrate that experiments are nicely consistent with a single component paired CF state. The assignment of the FQHE state with a single component $p$-wave state is also in line with another theoretical study by Zhu {\it et al.}~\cite{Zhu16}. Experimentally, the observation of the standard Jain sequences $\nu=n/(2n{\pm}1)$ on either side of the $1/2$ FQHE points to single layer physics, and the measured Fermi wave vector also indicates a single component CFFS~\cite{Mueed15}. Very recently, Singh {\it et al.}~\cite{Singh23} have observed, concurrent with the 1/2 FQHE state, anomalously strong FQHE states at 8/17 and 7/13, which are consistent with the theoretically predicted 
daughter fractions of the single-component Pf phase~\cite{Levin09a}.

\underline{$\nu=1/4$ in wide QWs:} There is evidence for FQHE at $\nu=1/4$ in wide QWs~\cite{Shabani09a,Shabani09b,Shabani13}. We show below that here the CFFS yields to $f$-wave pairing of CFs with increasing width or density. This is consistent with the calculation of Faugno {\it et al}.\cite{Faugno19} who found that the $22111$ parton state, which also represents $f$-wave pairing of CFs, has lower energy than the CFFS for sufficiently large QW widths.  

\underline{Bosons at $\nu_{b}=1$ and $\nu_{b}=1/3$:} Bosons in the LLL turn into CFs by attaching an odd number ($s$) of vortices to show FQHE at the Jain fractions $\nu_b=n/(sn\pm 1)$. One would expect a CFFS at $\nu_b=1$ by analogy to the electron problem, but, for the contact interaction, exact diagonalization (ED) studies show the MR-Pf to be energetically favorable. We refer the reader to Refs.~\cite{Cooper99,Regnault03,Chang05b,Korslund06,Cooper08,Viefers08} for details. In this article we show that the CFFS at $\nu_b=1$  is unstable to $p$-wave pairing for both the contact and the long range Coulomb interactions. In contrast, for $\nu_b=1/3$, we do not find any pairing instability of the CFFS for the Coulomb interaction. These results are consistent with ED studies~\cite{Liu20}. 

In summary, we find that the CF-BCS approach is reasonably successful in uncovering pairing instabilities in a number of different contexts. In particular, an accumulation of the experimental results and the present theoretical work makes a strong case for a single-component FQHE at $\nu=1/2$ and $1/4$ with pairing symmetries of $p$ and $f$ wave, respectively.

During the course of this work, we have found that 
the model for electron-background and background-background interaction affects the thermodynamic extrapolations obtained from trial wave functions in the spherical geometry. See Appendix ~\ref{appx-sphere} for a discussion of this issue. There and in Sec.~\ref{two-component} we present more accurate calculations for the phase diagram in the spherical geometry for $\nu=1/4$ and $\nu=1/2$. At $\nu=1/4$ an instability into the $22111$ parton state is still seen, but the phase boundary we obtain is somewhat different from that in Ref.~\cite{Faugno19}. At $\nu=1/2$, the MR Pfaffian state is found to have higher energy than the CFFS in the entire range of width and density studied, in contrast to the claim in Ref.~\cite{Zhao21}.

The plan for the remainder of the paper is as follows. In section \ref{sec:review}, we provide a brief review, for completeness, of the basics of CFs on a torus and also introduce certain known wave functions. In section \ref{sec:BCS}, we construct our CF-BCS wave function for spin-polarized CFs and show that the modified JK projection scheme produces wave functions that satisfy the proper quasi-periodic boundary conditions. In Sec.~\ref{sec:1over2} we study the nature of the state at $\nu=1/2$ in a wide QW and find that a $p$-wave pairing instability occurs as the QW width or the density is increased.  The theoretical phase diagram is in excellent agreement with the experimental one. We also find that asymmetry of the QW favors the CFFS state, and a two-component candidate state is energetically less favorable. In Sec.~\ref{sec:1over4}, we present the results of our calculations for $\nu=1/4$, which demonstrate an $f$-wave pairing instability as the QW width or the density is increased. A system of bosons confined to the LLL is considered in Sec.~\ref{sec:bosonic}. A $p$-wave pairing instability is seen at filling factor $\nu_b=1$ for both the contact and the Coulomb interactions. We provide a rough estimation for the gap from the condensation energy in Sec.~\ref{sec: gaps_from_Ec}. The paper is concluded in Sec.~\ref{sec:conclusions}. 
Many relevant details are presented in appendices.

\section{Composite fermions on a torus}
\label{sec:review}

For completeness, this section contains a review of various relevant wave functions in the torus geometry. A torus can be mapped to a parallelogram with periodic boundary conditions. We denote the two edges of the parallelogram by $L_1=L$ and $L_2=L\tau$, where $\tau=\tau_1+i\tau_2$ is a complex parameter. The modular parameter $\tau$ specifies the torus \cite{Gunning62}. We consider $L_1$ to be along the real axis. The magnetic field $\vec{B}=-B\hat{z}$ is perpendicular to the parallelogram. The positions of the particles are represented by the complex coordinates $z=x+iy$. We consider the symmetric gauge for our calculations, given by $A=\frac{B}{2}(y,-x,0)$. The single particle wave functions on the torus satisfy the periodic boundary conditions in the two directions:
\begin{eqnarray}
t(L_1)\psi(z,\bar{z}) =  e^{i\phi_1 } \psi(z,\bar{z})\\ \nonumber
t(L_2)\psi(z,\bar{z}) =  e^{i\phi_2 } \psi(z,\bar{z})
\end{eqnarray}
where $t(L_i)$ is the magnetic translation operator in the $L_i$ direction defined by
\begin{equation}
t(\xi) =  e^{-\frac{i}{2\ell ^2}\hat{\vec{z}}.(\vec{\xi} \times \vec{r})} T(\xi)
\end{equation}
with $\ell=\sqrt{\hbar c/eB}$ as the magnetic length and $T(\xi)$ as the translation operator for a vector $\xi$. The translation operator for a vector $\xi$ can be written as 
\begin{equation}
T(\xi) = e^{\xi \partial _z + \bar{\xi} \partial _{\bar{z}}}
\end{equation}
with $\xi = \xi _x + i \xi _y$. Some of the important relations which are needed to show the boundary conditions are
\beq
t(L_1)e^{z^2-|z|^2 \over 4 \ell ^2} = e^{z^2-|z|^2 \over 4 \ell ^2} T(L_1)
\eeq
\beq
t(L_2)e^{z^2-|z|^2 \over 4 \ell ^2} = e^{z^2-|z|^2 \over 4 \ell ^2}e^{-i\pi N_{\phi}(2z/L+\tau)} T(L_2)
\label{eq:L2}
\eeq
The many particle wave function must satisfy the property
\ba
t_j(L_1)\Psi(\{z_i\},\{\bar{z}_i\})= e^{i\phi_1}\Psi(\{z_i\},\{\bar{z}_i\})  \\ \nonumber
t_j(L_2)\Psi(\{z_i\},\{\bar{z}_i\})= e^{i\phi_2}\Psi(\{z_i\},\{\bar{z}_i\})
\label{eq:pbc}
\ea
where $t_j$ is the magnetic translation operator for the $j$th particle. 
\\

Below are some candidate states  at $\nu=1/2m$. The wave functions are confined to the  lowest Landau level (LLL)  and we do not include Landau level mixing in our calculations. 

{\it CFFS wave function:}
The CFFS wave function on a torus is given by:
\beq
\Psi^{\rm CFFS}_{1/2m,k_{\rm CM}} =  \mathcal{P}_{\rm LLL} {\rm Det}[e^{i\vec{k}_n\cdot\vec{r}_m}]\Psi_{1/2m,k_{\rm CM}}^{\rm L} 
\label{CFFS1}
\eeq
where $\mathcal{P}_{\rm LLL}$ is the LLL projection operator \cite{Pu18,Pu20b}. The allowed values of the wave vectors $\vec{k}$ are given by
\beq
\vec{k}_n = \left[n_1+{\phi_1\over 2\pi} \right]\vec{b_1} + \left[n_2+{\phi_2\over 2\pi} \right] \vec{b_2}
\label{eq:kn}
\eeq
where
\be
\vec{b}_1=\left({2\pi\over L},-{2\pi\tau_1\over L\tau_2}\right),\;\;
\vec{b}_2=\left(0,{2\pi\over L\tau_2}\right).
\ee
The discrete values of $\vec{k}$ are constrained by periodic boundary conditions. In what follows, we will also use the notation $k=k_x+ik_y$. In Eq.~\eqref{CFFS1}, $\Psi_{1/2m,k_{\rm CM}}^{\rm L} $ is the Laughlin wave function at filling fraction $\nu=1/2m$, which is written as
\begin{widetext}
\be
\label{Laughlin}
\Psi^{\rm L}_{1/m,k_{\rm CM}}[z_i,\bar{z}_i]=e^{\sum_i {z_i^2-\abs{z_i}^2 \over 4 \ell^2}}
\left[\elliptic{{\phi_1\over 2\pi m}+{k_{\rm CM}\over m}+{N-1\over2}}{-{\phi_2\over 2\pi}+{m(N-1)\over2}}{mZ \over L_1}{m\tau}\right] 
\prod_{i<j} \left[ \elliptic{{\frac12}}{{\frac12}}{z_{i}-z_{j}\over L_1}{\tau} \right]^{2m}
\ee
\end{widetext}
where $k_{\rm CM}$ takes values $k_{\rm CM}=0,1,\cdots 2m-1$; it selects the center of mass momentum sector. For our calculations, we select $k_{\rm CM}=0$.

We use the Jacobi theta functions with rational characteristics $\elliptic{a}{b}{z}{\tau}= \sum_{n=-\infty}^{\infty}e^{i\pi \left(n+a\right)^2\tau}e^{i2\pi \left(n+a\right)\left(z+b\right)}$\cite{Mumford07}, whose properties are summarized in Ref.~\cite{Sharma23}.

{\it MR wave function:} The MR wave function in disk geometry at $\nu=\frac{1}{2m}$ is given by:
\be
\Psi_{\rm MR}=\exp[-\sum_j|z_j|^2/4\ell^2]{\rm Pf}\left( {1\over z_j-z_k} \right) \prod_{j<k}(z_j-z_k)^{2m} \label{Pfdisk}
\ee
where Pf represents Pfaffian, which is defined, for an $N\times N$ (with even $N$) antisymmetric matrix $M_{ij}$, as
\beq
{\rm Pf}\{M_{ij}\}=\frac{1}{2^{N/2}(N/2)!}\sum_{\sigma}\prod_{i=1}^{N/2}M_{\sigma(2i-1)\sigma(2i)}
\label{eq:Pf}
\eeq
where $\sigma$ labels all permutations. The factor ${\rm Pf}\left( {1\over z_j-z_k} \right)$ represents a $p$-wave paired wave function for electrons, while $\prod_{j<k}(z_j-z_k)^{2m}$ attaches $2m$ vortices to electrons to convert the wave function into a paired state of CFs.  
On a torus, the MR wave function takes the form~\cite{Greiter92a,Chung07,Read96}
\begin{widetext}
\begin{equation}
\label{MR}
\Psi^{(a,b,k_{\rm CM})}_{{\rm MR}-p} = e^{\sum_i \frac{z_i^2 -|z_i|^2}{4 \ell^2}} \vartheta  \begin{bmatrix}
{\frac{\phi_1}{4\pi 2m }+\frac{k_{\rm CM}}{2m}+ \frac{(N-1)}{2}+\frac{(1-2a)}{4}} \\ {-\frac{\phi_2}{2\pi } + m(N-1)-\frac{(1-2b)}{2}}
\end{bmatrix}\Bigg({2mZ \over L} \Bigg |2m \tau \Bigg) 
\rm Pf\Bigg(\frac{\vartheta \begin{bmatrix}
{a} \\ {b}
\end{bmatrix}
\Bigg( {z_i-z_j\over L}\Bigg | \tau \Bigg )}{\vartheta 
\begin{bmatrix}
{1\over 2} \\ {1\over 2}
\end{bmatrix}
\Bigg( {z_i-z_j\over L}\Bigg | \tau \Bigg )}\Bigg) \prod_{i<j} \Bigg [ \vartheta 
\begin{bmatrix}
{1\over 2} \\ {1\over 2}
\end{bmatrix}
\Bigg( {z_i-z_j\over L}\Bigg | \tau \Bigg ) \Bigg ]^{2 \it m}
\end{equation}
The above wave function represents a $p$-wave paired state of CFs. Similarly, we write a MR type Pfaffian wave function with $f$-wave pairing at $\nu=1/2m$ (for $m \geq$ 2):
\begin{equation}
\label{MR(l=3)}
\Psi^{(a,b,k_{\rm CM})}_{{\rm MR}-f} = e^{\sum_i \frac{z_i^2 -|z_i|^2}{4 \ell^2}} \vartheta  \begin{bmatrix}
{\frac{\phi_1}{4\pi m }+\frac{k_{\rm CM}}{2m}+ \frac{(N-1)}{2}+\frac{(1-2a)}{4}} \\ {-\frac{\phi_2}{2\pi } + m(N-1)-\frac{(1-2b)}{2}}
\end{bmatrix}\Bigg({2mZ \over L} \Bigg |2m \tau \Bigg) 
 {\rm Pf}\Bigg( \Bigg [\frac{\vartheta \begin{bmatrix}
{a} \\ {b}
\end{bmatrix}
\Bigg( {z_i-z_j\over L}\Bigg | \tau \Bigg )}{\vartheta 
\begin{bmatrix}
{1\over 2} \\ {1\over 2}
\end{bmatrix}
\Bigg( {z_i-z_j\over L}\Bigg | \tau \Bigg )}\Bigg]^3 \Bigg ) \prod_{i<j} \Bigg [ \vartheta 
\begin{bmatrix}
{1\over 2} \\ {1\over 2}
\end{bmatrix}
\Bigg( {z_i-z_j\over L}\Bigg | \tau \Bigg ) \Bigg ]^{2\it m}
\end{equation}.
\end{widetext}
The numerator inside the Pfaffian factor is required to produce the desired boundary conditions, and the 
 parameters $(a,b)$ can take values $(0,\frac{1}{2}),(\frac{1}{2},0)$ or $(0,0)$ (for which the theta function is even under exchange of particles), which indicates a topological ground state degeneracy of three~\cite{Chung07}. As the $i$th and $j$th particles approach one another, the quantity 
${\vartheta \begin{bmatrix}
{a} \\ {b}
\end{bmatrix}
\Bigg( {z_i-z_j\over L}\Bigg | \tau \Bigg )}$ vanishes as $z_i-z_j$ for $(a,b)=(1/2,1/2)$, whereas it does not vanish for $(a,b)=(0,\frac{1}{2}),(\frac{1}{2},0)$ or $(0,0)$, approaching a constant instead. 
(The above wave function for MR-$f$  is ill defined for $\nu=1/2$, as the Jastrow factor does not have enough powers to cancel the divergence of the Pfaffian when two particles approach one another.)

\section{CF-BCS wave function}
\label{sec:BCS}

In this section, following the approach outlined in Ref.~\cite{Sharma21}, we construct BCS wave functions for spin polarized CFs at $\nu=\frac{1}{2}$ and $\frac{1}{4}$ for a general pairing, including $p$ and $f$-wave pairings. We then describe the projection scheme for our calculations.

The real space form of BCS wave function for fixed number of electrons is given by \cite{Pierre99}
\begin{equation}
\Psi_{\rm BCS}(\vec{r}_1,...\vec{r}_N) = {\rm Pf}(g^{(l)}(\vec{r}_i-\vec{r}_j))
\end{equation} 
where $g^{(l)}(\vec{r}_i-\vec{r}_j)$ can be expanded as
\begin{equation}
g^{(l)}(\vec{r}_i-\vec{r}_j) = \sum_{\vec{k}}g^{(l)}_{\vec{k}} e^{i{\vec{k}}\cdot(\vec{r}_i-\vec{r}_j)}.
\end{equation}
Here each $\vec{k},\vec{-k}$ is occupied only once. For an odd pairing symmetry $l$, following the BCS theory, we have
\beq 
g^{(l)}_{\vec{k}}\equiv\frac{v_{\vec{k}}}{u_{\vec{k}}}= \frac{\epsilon_{\vec{k}} - \sqrt{\epsilon_{\vec{k}}^2 +|\Delta^{(l)}_{\vec{k}}|^2}}{\Delta^{(l)*}_{\vec{k}}}=-g^{(l)}_{-\vec{k}}.
\eeq
where $\epsilon_{\vec{k}}=\hbar ^2{|\vec{k}|^2}/2m-\hbar ^2{|\vec{k_F}|^2}/2m$ and $\Delta_{\vec{k}}$ is the gap function. The quantities $\vec{k_F}$ and $m$ represent the Fermi wave vector and mass of the electron, respectively. 
Analogously, the CF-BCS wave function at $\nu=1/2m$ can be constructed as
\beq
\Psi _{\frac{1}{2m}}^{\rm BCS} = \mathcal{P}_{\rm LLL} {\rm Pf}\left(\sum _{\vec{k}} g^{(l)}_{\vec{k}} e^{i\vec{k}\cdot\left(\vec{r}_i-\vec{r}_j\right)}\right) \Psi^{\rm L}_{1/2m}
\eeq
where the mass of electron $m$ is replaced by effective mass of CF $m^*$. The form of the gap function for the $l$ pairing channel is $\Delta^{(l)}_{\vec{k}}=\Delta |\vec{k}|e^{-il\theta}$, where $\theta$ is the relative angle between the $k_x$ and $k_y$ components of the wave vector $\vec{k}$. The pairing channel is defined by the eigenvalues of rotations in $\vec{k}$ with  $\Delta^{(l)}_{\vec{k}}$ as eigenfunctions \cite{Read00}. The parameters $l=1$ and $l=3$ correspond to $p$-wave and $f$-wave pairing, respectively. It signifies the number of copropagating Majorana edge modes, giving a central charge $c=1+l/2$ \cite{Dubail11}. 

\begin{widetext}
Following Refs.~\cite{Pu18,Pu20b,Sharma21}, the direct projected CF-BCS state at $\nu=1/2m$ is given by 
\begin{equation}
 \Psi_{\frac{1}{2m}}^{\rm BCS} = e^{\sum_i \frac{z_i^2 - |z_i|^2}{4\ell^2}}\elliptic{{\phi_1\over 4\pi m}+{N-1\over 2}}{-{\phi_2\over 2\pi}+m(N-1)}{{2mZ\over L_1}}{2m\tau}{\rm Pf}\left[\sum_ng^{(l)}_{\vec{k}_n}\hat{F}_n(z_i,z_j)\right]\prod_iJ_i^m
\end{equation}
where 
\begin{equation}
J_i=\prod_{r \neq i} \vartheta 
\begin{bmatrix}
{1\over 2} \\ {1\over 2}
\end{bmatrix}
\Bigg( {z_i-z_r \over L}\Bigg | \tau \Bigg ),
\end{equation}
\begin{equation}
\hat{F}_n(z_i,z_j)=e^{-\frac{k_n\ell^2}{2}(k_n+2\Bar{k}_n)}e^{\frac{i}{2}(z_i - z_j)(k_n+\Bar{k}_n)}e^{ik_n\ell^2 \partial _{z_i}} e^{-ik_n\ell^2 \partial _{z_j}}, 
\end{equation}
and $\vec{k}_n$ are given by Eq.~\ref{eq:kn}
The above form of the wave function is not amenable to calculations for large systems, because LLL projection can be performed only for rather small systems. We therefore follow the Jain-Kamilla (JK) projection method \cite{Jain97, Jain97b}. 
The basic idea is to bring some of the Jastrow factors inside the Pfaffian matrix and: then project each matrix element of the Pf separately
\begin{equation}
{\rm Pf}\left[\sum_ng^{(l)}_{\vec{k}_n}\hat{F}_n(z_i,z_j)\right]\prod_iJ_i^{m}\rightarrow 
{\rm Pf}\left[\sum_ng^{(l)}_{\vec{k}_n}\hat{F}_n(z_i,z_j)J_i J_j\right](\prod_iJ^{m-1}_i).
\label{method1}
\end{equation} 
The right hand side is {\em not} a legitimate wave function in the torus geometry, because it does not satisfy the correct periodic boundary conditions. It was shown in Ref.~\cite{Sharma21}, following earlier work dealing with FQHE states\cite{Pu18,Pu20b}, 
that if we 
 replace the operators $e^{ik_n\ell^2 \partial _{z_i}} e^{-ik_n\ell^2 \partial _{z_j}} $ in $\hat{F}_n(z_i,z_j)$ by $e^{ik_n\ell^2 \hat{D}^{(j)} _{z_i}} e^{-ik_n\ell^2 \hat{D}^{(i)} _{z_j}}$, the boundary conditions are preserved.
Here the new derivative operator $\hat{D}^{(j)}_{z_i}$ is defined as
\be
\hat{D}^{(j)}_{z_i}\elliptic{1/2}{1/2}{z_i-z_l\over L}{\tau}\equiv
\begin{cases}
2{\partial\over \partial{z_i}}\elliptic{1/2}{1/2}{z_i-z_l\over L}{\tau} \quad {\rm if}\quad l=j\\
2m{\partial\over \partial{z_i}}\elliptic{1/2}{1/2}{z_i-z_l\over L}{\tau} \quad {\rm if}\quad l\neq j
\end{cases}
\ee
The final form of the JK projected wave BCS function is 
\begin{equation}
\label{BCS_paired_2}
    \Psi_{\frac{1}{2m}}^{\rm BCS} = e^{\sum_i \frac{z_i^2 - |z_i|^2}{4\ell^2}}\Bigg\{\vartheta
\begin{bmatrix}
{\phi_1\over 4\pi m}
 + {N-1 \over 2}\\ 
-{\phi_2\over 2\pi } + {m(N-1)}
\end{bmatrix}
\Bigg({2mZ \over L} \Bigg |2m \tau \Bigg) \Bigg \} {\rm Pf}(\tilde{M}_{ij})  \prod_{i<j}  \Bigg\{\vartheta \begin{bmatrix}
{1 \over 2} \\ {1 \over 2 }
\end{bmatrix}\Bigg(\frac{z_i - z_j}{L}\Bigg |\tau \Bigg) \Bigg\}^{2(m-1)}
\end{equation}
where the matrix element is:
\begin{eqnarray}
\label{JK2_Pf}
\tilde{M}_{ij}=\sum_{k_n}g^{(l)}_{\vec{k}_n}e^{-\frac{\ell^2}{2}k_n(k_n+2\Bar{k}_n)}e^{\frac{i}{2}(z_i-z_j)(k_n+\Bar{k}_n)} \prod_{\substack{m \\m \neq i,j}}   \vartheta  \begin{bmatrix}
{1 \over 2} \\ {1 \over 2 }
\end{bmatrix}\Bigg(\frac{z_i + i\textcolor{red}{2m}k_n\ell^2- z_m}{L}\Bigg |\tau \Bigg)  \nonumber \\ \prod_{\substack{n \\n \neq i,j}} \vartheta  \begin{bmatrix}
{1 \over 2} \\ {1 \over 2 }
\end{bmatrix}\Bigg(\frac{z_j - i\textcolor{red}{2m}k_n\ell^2- z_n}{L}\Bigg |\tau \Bigg)  \Bigg \{\vartheta  \begin{bmatrix}
{1 \over 2} \\ {1 \over 2 }
\end{bmatrix}\Bigg(\frac{z_i + i\textcolor{red}{2m}k_n\ell^2- z_j}{L}\Bigg |\tau \Bigg) \Bigg \}^2 
\end{eqnarray}
The JK projected wave function satisfies the periodic boundary conditions as shown in Appendix \ref{pbc}.
\end{widetext}

For  $\nu=1/4$ we have another choice for the JK projection, where we pull all of the Jastrow factor into the Pf to write  
 ${\rm Pf}\left[\sum_ng^{(l)}_{\vec{k}_n}\hat{F}_n(z_i,z_j)J^2_i J^2_j\right]$. We have tested this as well and found that the resulting wave function is very close to that in Eq.~\eqref{method1} and does not change any conclusions.

We define a dimensionless variational parameter $\tilde{\Delta}$ for the CF-BCS wave function \cite{Sharma23}:   
\beq 
\tilde{\Delta}={|\Delta^{(l)} _{k_F}| \over \hbar^2|k_F|^2/2m^*}.
\eeq
This is the gap parameter. We introduce another parameter, $k_{\rm cutoff}$, such that only wave vectors $|\vec{k}|\leq k_{\rm cutoff}$ participate in pairing. The quantity $g^{(l)}_{\vec{k_n}}$ can then be re-written as \cite{Sharma23} 
\beq
\label{gk2}
g^{(l)}_{\vec{k}_n} =
\begin{cases}
{\abs{k_n}^2 - \abs{k_F}^2-\sqrt{(\abs{k_n}^2 - \abs{k_F}^2)^2+\tilde{\Delta}^2|k_F|^2\abs{k_n}^2}\over \tilde{\Delta}|k_n|e^{il\theta}|k_F|} & \text{$|\vec{k}_n| \leq k_{\rm cutoff}$}  \nonumber \\
0 & \text{$|\vec{k}_n|>k_{\rm cutoff}$} 
\end{cases}
\\
\eeq
For our calculations, we determine the magnitude of $k_F$ using the relation:
\beq
\pi |k_F|^2 = N |\vec{b_1} \cross \vec{b_2}|
\eeq
We perform our numerical calculations for even number of particles with $N=12$ and $N=32$ particles. The Fermi sea configuration is approximately circular for these system sizes.

\section{Origin of FQHE at $\nu=1/2$ in wide quantum wells}
\label{sec:1over2}

\begin{figure}[t]
\centering
\includegraphics[width=\linewidth]{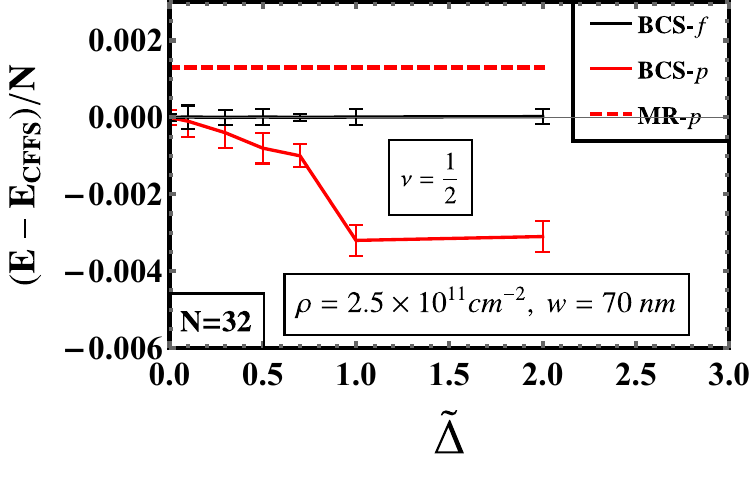} 
\caption{\label{variation-half}The energy per particle for various candidate states at $\nu=1/2$ as a function of $\tilde{\Delta}$ at density= $2.5 \times 10^{11}$cm$^{-2}$ and QW width=$70$nm for a system of 32 particles. The torus geometry is used for the calculation. The BCS-$p$ $(l=1)$ state clearly has lower energy than the CFFS. The energies are measured relative to the energy of the CFFS state ($\rm E_{CFFS }$) in units of $e^2/\epsilon \ell$. 
}
\end{figure}

\begin{figure}[t]
\centering
\includegraphics[width=\linewidth]{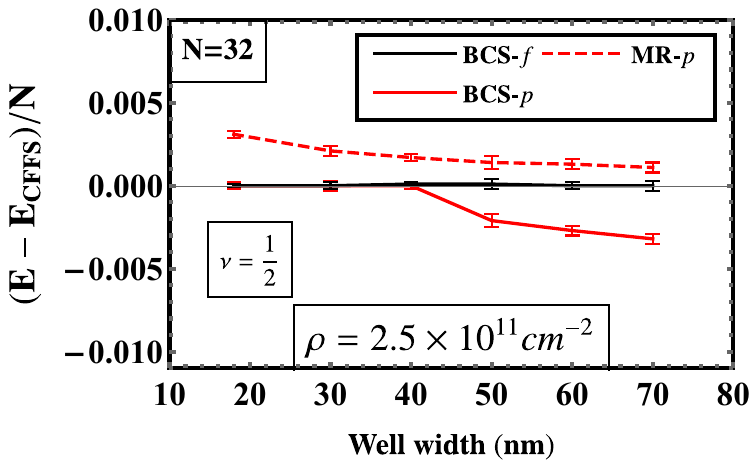} 
\includegraphics[width=\linewidth]{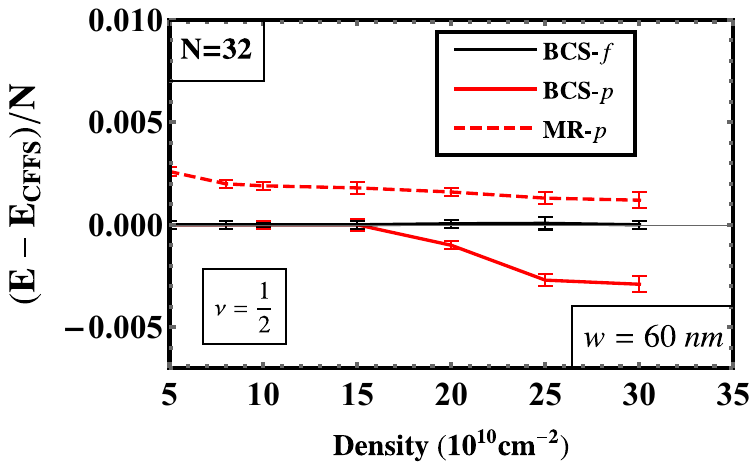} 
\caption{\label{energy-half}(upper panel) The energy as a function of the well width for several candidate states at density= $2.5 \times 10^{11}$cm$^{-2}$. (lower panel) Energy as a function of density for a fixed well width = 60 nm. The energies are plotted with respect to the energy of the CFFS, in units of $e^2/\epsilon \ell$. The calculations are performed for $N=32$ particles on a torus. 
}
\end{figure}

FQHE at $\nu=1/2$ was seen in wide QWs three decades ago~\cite{Suen92,Suen94,Luhman08,Shabani13}. Because the $1/2$ state in narrow QWs is undoubtedly a CFFS, experiments imply a phase transition into a FQHE state as a function of the QW width or the electron density.  There has been much discussion regarding the nature of the 1/2 FQHE, and especially on whether it is a single-component or a two-component state. In the next subsection, we show that theory predicts a CF pairing instability within a single component state residing within the lowest subband of the wide QW. In the subsequent section we present variational calculations indicating that a candidate two-component state has higher energy than the CFFS. In the last subsection, we search for a quantum phase transition in the nearby $\nu=6/13$ FQHE state as a function of the QW width and the electron density.

\subsection{BCS pairing instability in the lowest subband}

We begin by exploring the possibility of a pairing transition assuming a single component origin, that is, assuming that all electrons occupy only the lowest symmetric subband.  
The effect of finite QW width can be incorporated by modifying the Coulomb interaction as 
\begin{equation}
V_{\rm C}(\vec{r}_1-\vec{r}_2)=\int dw_1 \int dw_2\frac{|\xi(w_1)|^2| \xi(w_2)|^2}{\sqrt{|\vec{r}_1-\vec{r}_2|^2+(w_1 - w_2)^2}},
\label{eq:VC}
\end{equation}
 where $\xi(w)$ is the electron wave function in the transverse direction, $w$ is the corresponding coordinate in the transverse direction, and $r$ is the in-plane distance between the electrons. The transverse wave function $\xi(w)$ is obtained in a local density approximation (LDA)~\cite{Ortalano97, Martin20}. The coordinates are in units of the magnetic length $\ell=\sqrt{\hbar c/eB}$ and the energy is in units of $e^2/\epsilon \ell$. For the torus geometry, we use a periodic form of the interaction. The details of the calculations are given in Appendix \ref{appx-periodic-int}. 

\begin{figure}[h]
		\includegraphics[width=\linewidth]{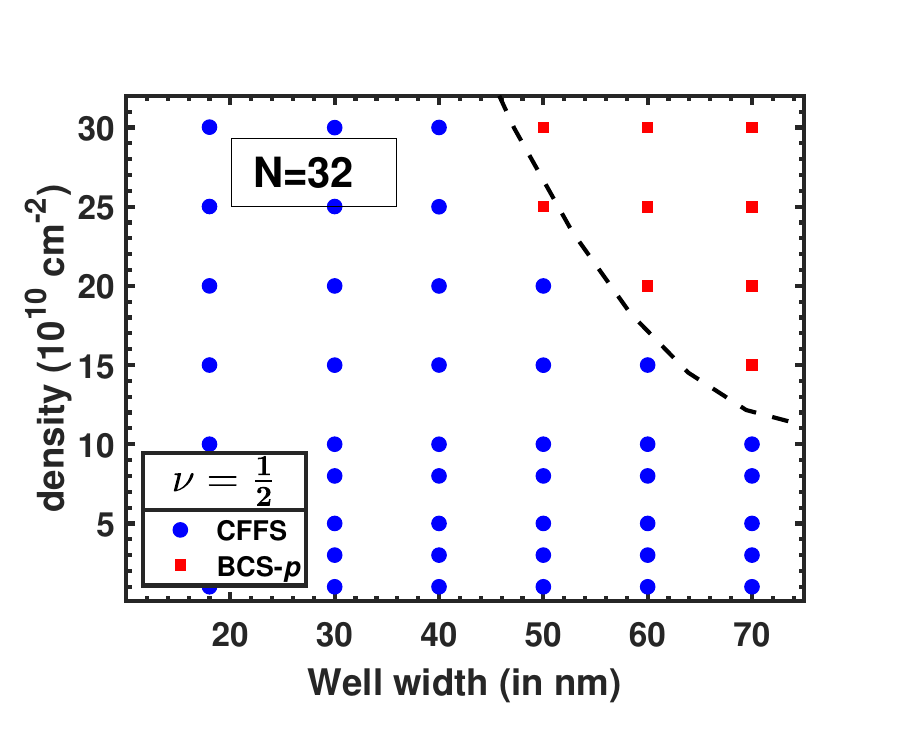}		
\caption{\label{Phase diagram half} This figure shows the minimum energy state at $\nu=1/2$ (blue dots for the CFFS and red squares for the $p$-wave paired state; all other candidate states have higher energies) as a function of the electron density and the QW width.  The calculation is performed for a system of 32 particles on a torus. The dashed line depicts the experimental phase boundary separating the compressible and the FQHE states at $\nu=\frac{1}{2}$ \cite{Suen92,Suen94,Luhman08,Shabani13}.
}
\end{figure}

In Fig.~\ref{variation-half}, we consider the system with a large density and QW width ($\rho = 2.5 \times 10^{11}$cm$^{-2}$ and $w=70~nm$) and plot the energies of the CFFS, BCS-$p$, BCS-$f$ and MR-$p$ states as a function of $\tilde{\Delta}$, where for each value of $\tilde{\Delta}$ we vary $k_{\rm cutoff}$ to find the lowest energy for the CF-BCS wave functions. (The energies of the CFFS and MR-$p$ states are independent of $\tilde{\Delta}$.) All energies here and below are quoted in units of $e^2/\epsilon \ell$.  The results indicate a $p$-wave pairing instability for these parameters.

Fig.~\ref{energy-half} shows the lowest energies of various paired states as a function of the QW width for a fixed electron density (upper panel) and also as a function of the density for a fixed QW width (lower panel). An instability occurs from the CFFS state into the BCS-$p$ state as either the QW width or the density is increased. We note that the minimum energy of the CF-BCS state for any $l$ is always less than or equal to that of the energy of the CFFS, because the CFFS is a special case of the CF-BCS states (obtained when $k_{\rm cutoff}=k_F$). The reader will notice that for many parameters all CF-BCS states have the same energies as the CFFS; there is no pairing instability for these parameters.

We obtain the phase diagram for a 32 particle system at $\nu=1/2$, shown in  Fig.~\ref{Phase diagram half}. A blue dot indicates that the CFFS has the lowest energy for that QW width and density, whereas the red squares mark parameters where the BCS-$p$ has the lowest energy. The dashed line is the  phase boundary obtained from experiments~\cite{Suen92,Suen94,Luhman08,Shabani13}. Clearly, the theoretical phase boundary is in excellent agreement with the experimental one.

The CFFS is known to be an excellent variational state at $\nu=1/2$~\cite{Rezayi00, Balram16b, Pu18, Zhao21}. It is therefore significant and nontrivial that a paired state has been found that has a lower energy than the CFFS. The 
rather small energy gains of 0.002 - 0.003 $e^2/\epsilon l$ per particle due to pairing (Fig.~\ref{energy-half}) indicate the quantitative accuracy required to capture this physics. Note that the MR-$p$ state has higher energy than the CFFS for all parameters considered in our study. (The competition between these two states in the spherical geometry is discussed in Appendix \ref{appx-sphere}.) That underscores the importance of the CF-BCS formalism in the explanation of the experiments.

Finally, we consider the role of asymmetry. Experiments have found that the incompressibility at $\nu=1/2$ is lost as the QW is made sufficiently asymmetric~\cite{Shabani09b}. This has been taken as evidence for two-component nature of the FQHE state. However, a similar effect may occur even for a single component FQHE, because making the transverse wave function asymmetric alters the interaction between electrons occupying the lowest symmetric subband. Indeed, in the limit of very large asymmetry, when all of the wave function is confined to one half of the quantum well, we end up with a narrow QW and expect a CFFS rather than a FQHE state. 

\begin{figure}[t]
\centering
\includegraphics[width=\linewidth]{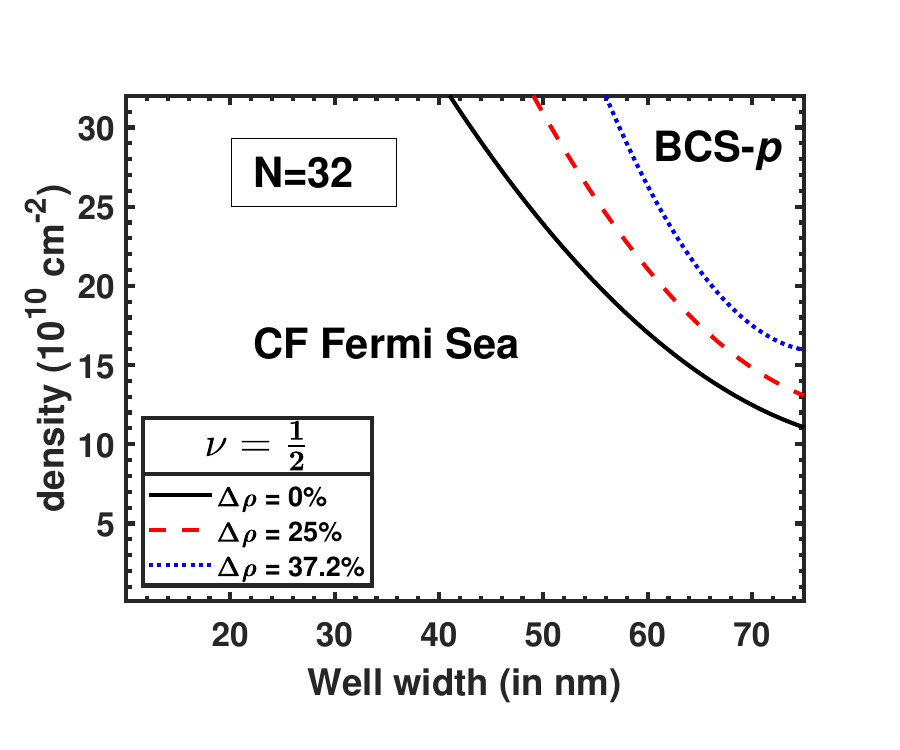} 
\caption{\label{asymmetry-half} This figure shows the variation in the phase boundary separating the CFFS and the CF $p$-wave paired state at $\nu=1/2$ as a function of the degree of asymmetry of the charge distribution in the QW (defined in the text). 
}
\end{figure}

We have investigated the effect of asymmetry of quantum well on the phase diagram.  We introduce asymmetry by placing the dopant layers at different distances on two sides of the QW in the self-consistent LDA calculation. We take as a measure of the asymmetry the quantity 
$\Delta \rho = (\rho_L-\rho_R)/(\rho_L+\rho_R)$, 
where $\rho_L$ and $\rho_R$ denote the densities in the left half and the right half of the QW, with the total density given by $\rho_L+\rho_R$. In Fig.~\ref{asymmetry-half}, we plot the theoretical phase boundaries for three different values of $\Delta \rho$. We find that the phase boundary shifts towards larger QW width or larger density as the QW is made asymmetric. This is consistent with the experimental observation that the resistance minimum at $\nu=1/2$ becomes weaker with increasing asymmetry~\cite{Shabani09b}.

\subsection{Candidate two-component state}
\label{two-component}

Two-component states may be relevant for wide QWs because, with increasing QW width, the energy separation between the lowest symmetric and antisymmetric subbands, denoted $\Delta_{\rm SAS}$, becomes small and the system becomes more and more like a bilayer. For an ideal bilayer, namely the system of two two-dimensional layers separated by a distance $d$, one expects two independent $1/4$ CFFSs in the limit of large $d$ and a pseudospin singlet CFFs at small $d$~\cite{Faugno20}. At intermediate $2\lessapprox d/\ell \lessapprox 3$, the Halperin 331 state~\cite{Halperin83} is a better variational state than these two~\cite{Scarola01b} and has been observed in a bilayer system~\cite{Eisenstein92}. One may ask if the Halperin 331 state is also relevant for the wide QWs of interest here. Its wave function in the disk geometry is given by 
\be
\Psi_{\rm 331}=\prod_{j<k}(z_j-z_k)^3 \prod_{j<k}(z_{[j]}-z_{[k]})^3 \prod_{j,k}(z_j-z_{[k]})
\label{eq:331}
\ee
where $j, k = 1, \cdots, N/2$, $[j]=N/2+j$, $z_j$ denotes the electron position within the two-dimensional plane, and we have suppressed the Gaussian factors for simplicity. For an ideal bilayer, the interaction between electrons in the same layer has the usual form, $e^2/\epsilon |z_j-z_k|$ and $e^2/\epsilon |z_{[j]}-z_{[k]}|$, whereas the interaction between electrons in different layers is given by $e^2/\epsilon \sqrt{|z_j-z_{[k]}|^2+d^2}$. 

For the present problem, we can either  define the right and the left basis functions as $\xi_L(w)=(\xi_S(w) +\xi_A(w))/\sqrt{2}$ and $\xi_R(w)=(\xi_S(w) -\xi_A(w))/\sqrt{2}$, where $\xi_S(w)$ and $\xi_A(w)$ are the transverse wave functions of the lowest symmetric and antisymmetric subbands obtained in LDA. Equivalently, we can consider the symmetric and antisymmetric subbands as the two components.  We will perform the calculations in Haldane's spherical geometry~\cite{Haldane83}, where, as discussed in Appendix \ref{appx-sphere}, it is important to be careful about the electron-background and background-background contributions to the energy. As discussed in that Appendix, the most accurate extrapolations to the thermodynamic limit are obtained when the background charge is assumed to have the same transverse charge distribution as the electrons, which is what we will do.  In addition to the Coulomb energy, another contribution to the energy is given by $\Delta_{\rm SAS}/2$ per particle, where an estimate for $\Delta_{\rm SAS}$ can be obtained from the LDA calculation.

For a single component state, we write the interaction term including the electron-electron, electron-background, and background-background interaction, as 
\be
\hat{V}={1\over 2}\int d\vec{R} \int d\vec{R}' [\hat{\rho}(\vec{R}) - \rho(\vec{R})] V(\vec{R}-\vec{R}') [\hat{\rho}(\vec{R}') - \rho(\vec{R}')] \nonumber
\ee
where $\vec{R}=(\vec{r},w)$ is a three-dimensional coordinate, $V(\vec{R})$ is the Coulomb interaction $e^2/\epsilon|\vec{R}|$, $\hat{\rho}(\vec{R})$ is the electron density operator and $\rho(\vec{R})=\langle \hat{\rho}(\vec{R})\rangle$. The expectation value $\langle \hat{V} \rangle$ is given by 
\begin{eqnarray}
&& {1\over 2} \int d\vec{R} \int d\vec{R}' \langle \hat{\rho}(\vec{R}) V(\vec{R}-\vec{R}') \hat{\rho}(\vec{R}') \rangle \nonumber \\
&&- {1\over 2} \int d\vec{R} \int d\vec{R}' {\rho}(\vec{R}) V(\vec{R}-\vec{R}'){\rho}(\vec{R}').
\end{eqnarray}
The last term is the electron-background and background-background contribution, denoted below by $E_b$. 
By integrating over the transverse coordinates, one can rewrite it  as 
\be
E_b=- {1\over 2}\int d\vec{r} \int d\vec{r}' {\rho}(\vec{r}) V_{\rm C}(\vec{r}-\vec{r}'){\rho}(\vec{r}') 
\ee
where the effective interaction $V_{\rm C}$ is given in Eq.~\eqref{eq:VC}.

For a two component state, we write the interaction as (with the components labeled $S$ and $A$):
 \begin{eqnarray}
&& {1\over 2}\int d\vec{R} \int d\vec{R}' [\hat{\rho}_S(\vec{R}) - \rho_S(\vec{R})] V(\vec{R}-\vec{R}') [\hat{\rho}_S(\vec{R}') - \rho_S(\vec{R}')]\nonumber \\
&& +{1\over 2}\int d\vec{R} \int d\vec{R}' [\hat{\rho}_A(\vec{R}) - \rho_A(\vec{R})] V(\vec{R}-\vec{R}') [\hat{\rho}_A(\vec{R}') - \rho_A(\vec{R}')]\nonumber \\
&&+ \int d\vec{R} \int d\vec{R}' [\hat{\rho}_S(\vec{R}) - \rho_S(\vec{R})] V(\vec{R}-\vec{R}') [\hat{\rho}_A(\vec{R}') - \rho_A(\vec{R}')] . \nonumber \\
\end{eqnarray}
Now the sum of the electron-background and background-background contribution is given by
\be \label{eq:2component}
E_{b}=- {1\over 2} \int d\vec{R} \int d\vec{R}' {\rho}(\vec{R}) V(\vec{R}-\vec{R}'){\rho}(\vec{R}'), 
\ee
where $\rho(\vec{R})=\rho_{S}(\vec{R})+\rho_{A}(\vec{R})$. This can be rewritten as a two-dimensional problem by introducing the effective interaction
\begin{equation}
V_{\rm C,\sigma\sigma'}(\vec{r}-\vec{r}')={e^2\over \epsilon}\int dw \int dw'\frac{|\xi_\sigma(w)|^2| \xi_{\sigma'}(w')|^2}{\sqrt{|\vec{r}-\vec{r}'|^2+(w - w')^2}},
\end{equation}
where $(\sigma,\sigma') = (S,S), (A,A), (S,A), (A,S)$ [evidently, we have $V_{\rm C,AS}=V_{\rm C,SA}$] so that
\be \label{eq:2component}
E_{b}=- \sum_{\substack{\sigma=A,S \\ \sigma'=A,S}} {1\over 2} \int d\vec{r} \int d\vec{r}' {\rho_{\sigma}}(\vec{r}) V_{\rm C,\sigma\sigma'}(\vec{r}-\vec{r}'){\rho_{\sigma'}}(\vec{r}'),
\ee
where $\rho_{\sigma}(\vec{r})$ is the $\sigma$ component's two-dimensional electron density.

\begin{figure}[b]
		\includegraphics[width=\linewidth]{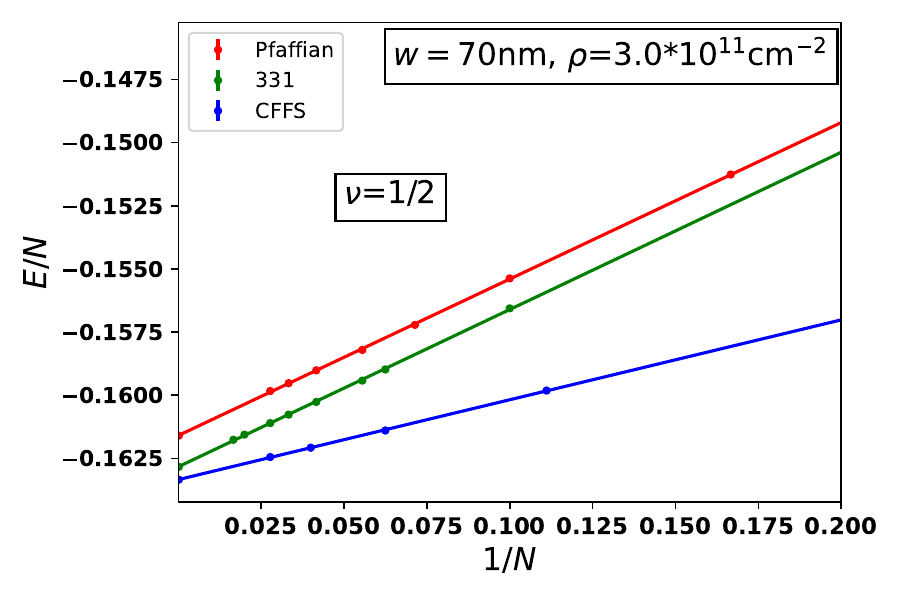}		
\caption{\label{sphere-one-half-therm} The energy $E/N$ of the CFFS, Pfaffian and the 331 states as a function of $1/N$ for QW width $w=70$ nm and electron density $\rho=3.0 \times 10^{11}$ cm$^{-2}$ at $\nu=1/2$.  The spherical geometry is used for the calculation. The energies are plotted in units of $e^2/\epsilon \ell$. The energy of the 331 state does not includes the contribution ${1\over 2}\Delta_{\rm SAS}$ per particle.
}
\end{figure}

We have obtained the thermodynamic limits for the energies of the Halperin 331, Pf and CFFS states in the spherical geometry for a range of QW widths and densities.  Some details are given in Appendix \ref{appx-sphere}. We find that the CFFS has lower energy than the 331 and the Pf states in the entire range of parameters shown in Fig.~\ref{Phase diagram half}.  Fig.~\ref{sphere-one-half-therm} shows the thermodynamic extrapolation for a 70 nm wide QW with density $3\times 10^{11}$ cm$^{-2}$. We note that the energy of the 331 state shown in this figure does not include the contribution $\Delta_{\rm SAS}/2$ (the LDA value is $\Delta_{\rm SAS}=5.65$ K for the  parameters in Fig.~\ref{sphere-one-half-therm}), which will lead to a further enhancement of its energy. Our calculations thus do not support the two-component 331 state. We note that this conclusion is based on a single wave function, and we have not considered a two-component CF-BCS type wave function.

\subsection{Daughter states near $\nu=1/2$}

Interestingly, the nature of the FQHE states in the vicinity of $\nu=1/2$ may also hold a clue into the origin of the FQHE at $\nu=1/2$. Let us recall the situation in the second LL where Kumar {\it et al.}\cite{Kumar10} found that the FQHE at $\nu=2+6/13$ is anomalously strong. Note that $6/13$ is routinely seen in the LLL; it belongs to the standard Jain sequence $n/(2n+1)$ and understood as six filled $\Lambda$ levels of CFs. However, the state at $2+6/13$ appears to have a different origin, as no FQHE is seen at $2+3/7$, $2+4/9$ and $2+5/11$.  The work by Balram {\it et al.}~\cite{Balram18,Balram18a} provides insight into the nature of this state by employing the parton theory of the FQHE~\cite{Jain89b}. Briefly, the parton construction proceeds by dividing each electron into an odd number ($m$) of fictitious fermions called partons, placing each parton species into an integer quantum Hall state, and finally identifying the partons to yield back the physical electrons. The wave function of the resulting state is $\Psi_{\nu}={\cal P}_{\rm LLL}\prod_{\lambda=1}^m\Phi_{n_\lambda}$, where $\Phi_n$ is the wave function of $n$ filled LLs, ${\cal P}_{\rm LLL}$ denotes projection into the LLL, and the filling factor is given by $\nu=[\sum_{\lambda=1}^m (n_\lambda)^{-1}]^{-1}$. This state is denoted as the $n_1 n_2\cdots n_m$ parton state. Negative values of $n$ are allowed; these are denoted by $\bar{n}$, with $\Phi_{-|n|}=[\Phi_{|n|}]^*$. The parton states $n11$ and $\bar{n}11$ represent the standard Jain CF states at $n/(2n+1)$ and $n/(2n-1)$. Ref.~\cite{Balram18} showed that the the $\bar{2}\bar{2}111$ state at $\nu=1/2$ lies in the anti-Pf phase and provides as good an account of the FQHE state at $2+1/2$ as the anti-Pf~\cite{Balram21b}, and the next state in this sequence, namely $\bar{3}\bar{2}111$, gives a satisfactory account of the FQHE state at $2+6/13$. The $\bar{3}\bar{2}111$ state has high overlap with the exact Coulomb state and also has lower energy than the Jain CF 6/13 state (i.e., the 611 parton state)~\cite{Balram18a, Balram20a}. This is consistent with the 5/2 FQHE being in the anti-Pf phase. (In the absence of LL mixing the Pf and the anti-Pf are equally plausible, but LL mixing is believed to select one of them.)

Levin and Halperin~\cite{Levin09a} have constructed a hierarchy emanating from the Pf and anti-Pf sates. They find that the first daughters of the Pf occur at 8/17 and 7/13, whereas those of the anti-Pf at 9/17 and 6/13. 
(The $\bar{3}\bar{2}111$ parton state is topologically equivalent to the Levin-Halperin daughter state at 6/13~\cite{Balram20a}. While the hole conjugate of $\bar{3}\bar{2}111$ provides a wave function at 7/13, the parton construction does not provide a simple wave function for 9/17 or 8/17.) 
Huang {\it et al.}~\cite{Huang22} have observed that the 1/2 states in bilayer graphene have either 8/17 and 7/13 or 9/17 and 6/13 flanking them, which enables the authors to deduce whether the 1/2 state is in the Pf or the anti-Pf phase. Very recent experiments by Singh {\it et al.}~\cite{Singh23} find that for a range of parameters the 1/2 state in wide QWs coexists with the Jain $n/(2n\pm 1)$ CF states, but as the 1/2 FQHE becomes stronger, there is a striking quantum phase transition at 8/17 and 7/13 into unusually strong FQHE states, which 
is consistent with the $\nu=1/2$ state being in the same phase as the Pf or the $p$-wave CF-BCS state.  

\begin{figure}[t]
\centering
\includegraphics[width=0.8\linewidth]{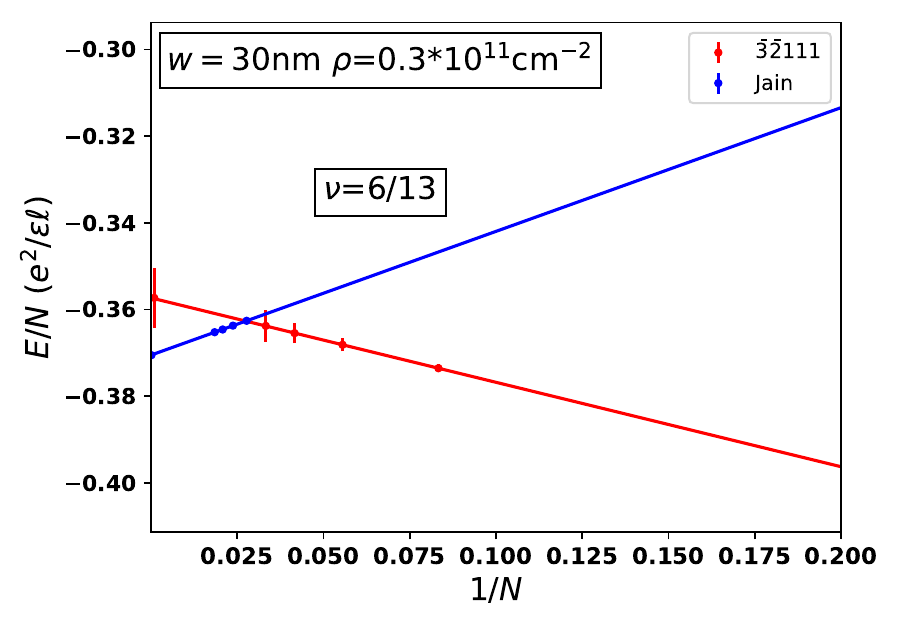} 
\includegraphics[width=0.8\linewidth]{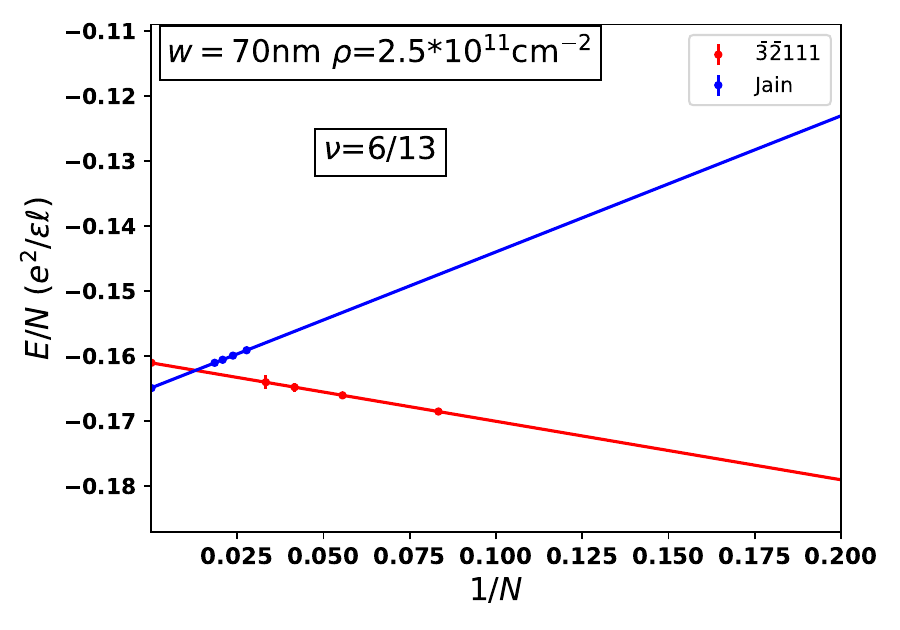} 
\includegraphics[width=0.8\linewidth]{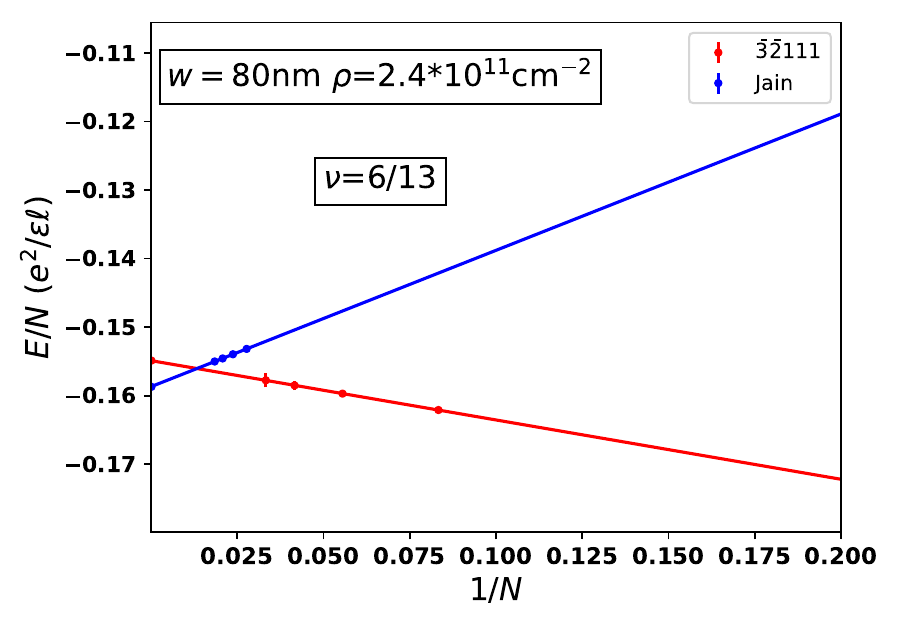} 
\caption{\label{fig:daughter} The energies of the Jain CF state and the $\bar{3}\bar{2}111$ parton state at $\nu=6/13$ as a function of $1/N$ for three different systems. In all cases, the former has the lower energy in the thermodynamic limit.
}
\end{figure}

We have investigated if theory can provide evidence of a phase transition at $\nu=6/13$ from the CF state to the $\bar{3}\bar{2}111$ parton state as the density and/or width of the QW is increased. (Note that because our theory does not include LL mixing, it is not able to distinguish between the Pf and the anti-Pf.)
We have calculated the energies of the two states for several parameters, and as shown in Fig.~\ref{fig:daughter}, the Jain CF state has lower energy in the thermodynamic limit. However, we stress that the comparison here is analogous to comparing the energies of the CFFS and the MR Pf states at $\nu=1/2$, which finds that the latter has higher energy and which thus fails to discover a pairing instability. A more accurate wave function for the paired state, namely the CF-BCS wave function is needed to capture, theoretically, the pairing instability of the CFFS. Given that the energies of the CF state and the $\bar{3}\bar{2}111$ parton state at $\nu=6/13$ are close, we believe that a slight improvement in the latter would be needed to reveal an instability of the Jain 6/13 state into the $\bar{3}\bar{2}111$ parton state. While we are unable to implement such an improvement at the moment, this brings out the subtlety of the physics in play here.

\section{CF Pairing at $\nu=1/4$ in wide quantum wells}
\label{sec:1over4}

\begin{figure}[t]
\centering
\includegraphics[width=\linewidth]{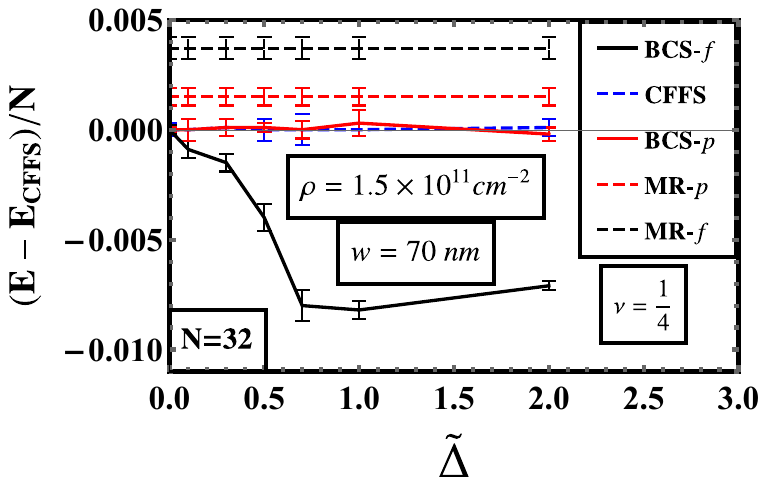} 
\caption{\label{variation}The energy per particle as a function of  $\tilde{\Delta}$ for several candidate states at $\nu=1/4$. The results are for the density= $1.5 \times 10^{11}$cm$^{-2}$ and QW width=$70$nm for a system of 32 particles. The torus geometry is used for the calculation. The BCS-$f$ $(l=3)$ state clearly has lower energy than the other states considered in our study. The energies are quoted relative to the energy of the CFFS ($\rm E_{CFFS}$) in units of $e^2/\epsilon \ell$. 
}
\end{figure}

\begin{figure}[t]
\centering
\includegraphics[width=\linewidth]{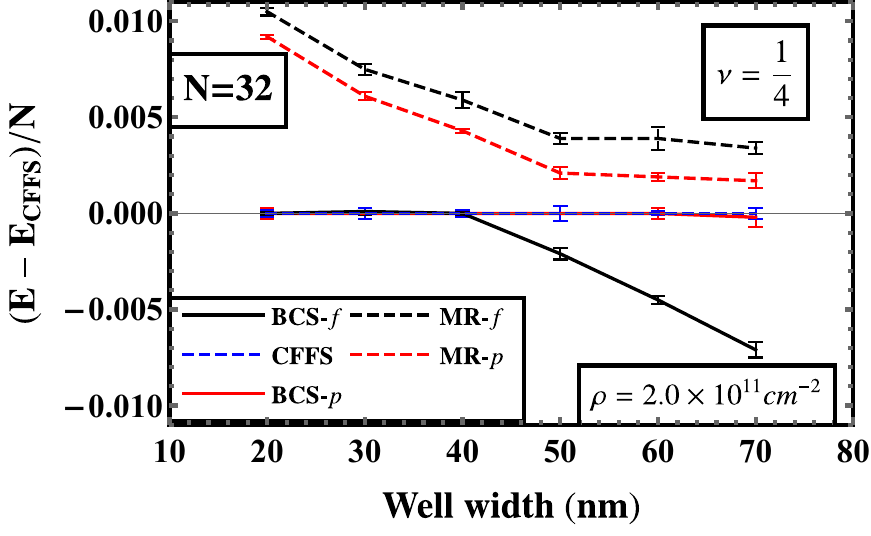} 
\includegraphics[width=\linewidth]{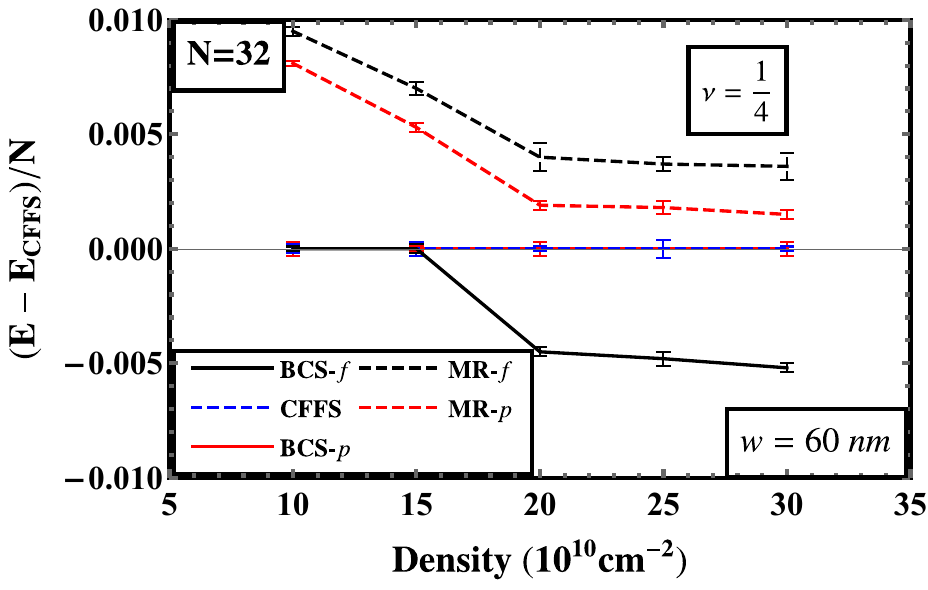}
\caption{\label{energy}(upper panel) Plot of per particle energy as a function of QW width at a fixed density of $20 \times 10^{10}$cm$^{-2}$ at $\nu=1/4$.  (lower panel) Plot of energy as a function of density for a fixed QW width=$60$nm at $\nu=1/4$. The energies are plotted with respect to the energy of the CFFS ($\rm E_{CFFS}$)  state in units of $e^2/\epsilon \ell$. The calculations are performed for 32 particles on a torus. 
}
\end{figure}

\begin{figure}[h]
		\includegraphics[width=\linewidth]{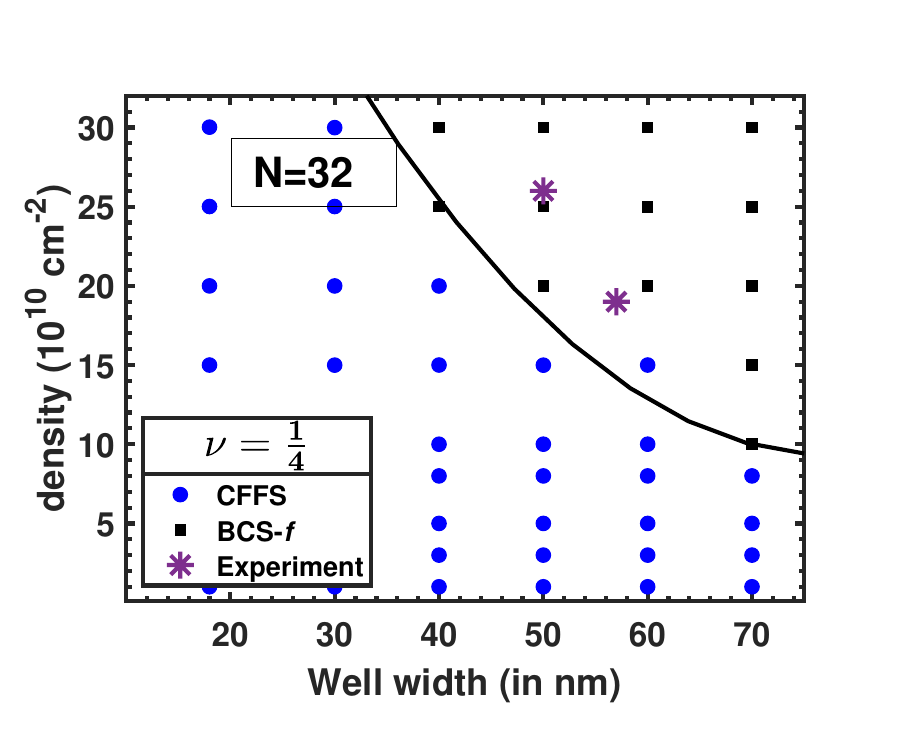}
		
\caption{\label{Phase diagram} This figure indicates the minimum energy state at $\nu=1/4$ (blue dots for the CFFS state and black squares for the $f$-wave paired state; all other candidate states have higher energies) as a function of the density and the QW width for a system of 32 particles. The torus geometry has been used for the calculation.  The solid line represents the approximate theoretical phase boundary. The stars mark  the experimental onset of the FQHE as a function of the density~\cite{Shabani09a}. 
}
\end{figure}

The $1/4$ state in narrow QWs  is well confirmed, both experimentally~\cite{Pan99,Hossain19} and theoretically~\cite{Liu20}, to be a CFFS of composite fermions carrying four flux quanta.  There is evidence for FQHE at $\nu=1/4$ in wide QWs \cite{Luhman08,Shabani09a}, again suggesting a pairing instability as a function of the QW width. Ref.~\cite{Faugno19} studied several wave functions and found that the 22111-parton wave function, which signifies an $f$-wave pairing of composite fermions, has the lowest energy of all wave functions considered, and in particular has lower energy than the CFFS and the Pfaffian states, for large QW widths. We revisit this issue within the CF-BCS approach.

Fig.~\ref{variation} shows the energy as a function of $\tilde{\Delta}$ for quantum QW width of 70 nm and density of $1.5 \times 10^{11}$cm$^{-2}$, where the energy at each $\tilde{\Delta}$ is obtained by varying $k_{\rm cutoff}$.  The $f$-wave CF-BCS state has the lowest energy. 

Fig.~\ref{energy} shows the lowest energy as a function of the QW width for a fixed density (upper panel), and the also as a function of the density for a fixed QW width (lower panel). A pairing instability occurs as either the QW width or the density is increased. Fig.~\ref{Phase diagram} depicts the phase diagram determined for a system of 32 particles as a function of the QW width and the electron density. At low densities and small QW widths, the CFFS is stabilized, while for large densities and large QW widths, the BCS-$f$ state minimizes the energy. The stars mark densities where FQHE begins to be seen in experiments for two different QW widths \cite{Shabani09a}. These are in decent agreement with the theoretical phase boundary, thus supporting $f$-wave pairing as the mechanism of the observed FQHE state. It is noted that the MR-$f$ and MR-$p$ wave function has a higher energy than the CFFS state in the entire parameter range studied here.

\section{Composite-fermion pairing for bosons at $\nu_b=1,\;1/3$}
\label{sec:bosonic}

\begin{figure}[h]
		\includegraphics[width=\linewidth]{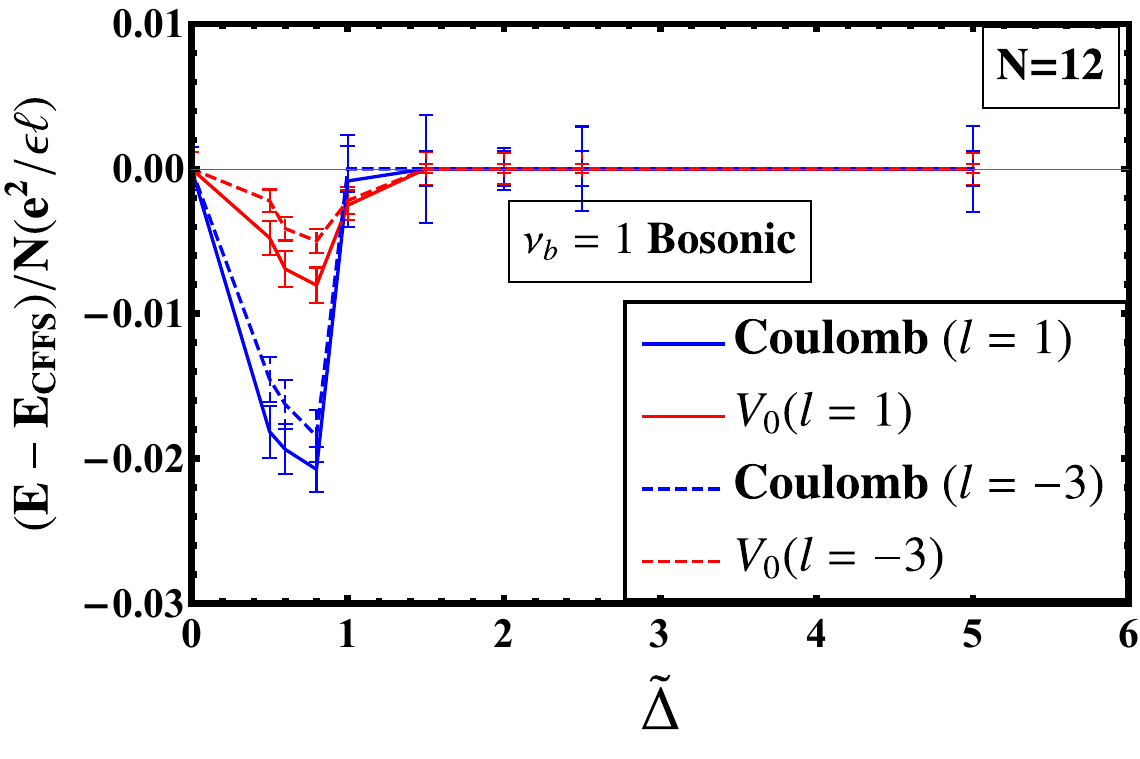}
		\caption{\label{fig:Bosonic-one} The energy per particle for the bosonic BCS state with $l=1$ pairing at $\nu_b=1$ for two interactions: the Coulomb interaction and the contact interaction. The results are for a system of 12 particles on a torus. The energies are quoted relative to the energy of the bosonic CFFS state in units of $e^2/\epsilon \ell$. The minimum energy state occurs at $\tilde{\Delta}\approx 0.8$. 
}
\end{figure}

We finally ask what the CF-BCS formalism predicts for bosons in the LLL.  Bosons in the LLL can bind to odd number ($m$) of flux quanta to form composite fermions, which can fill an integer number of CF-LLs to produce FQHE states at fillings $\nu_b=n/(mn\pm 1)$. The Jain CF wave functions for the bosonic FQHE states at $\nu=n/(n+1)$ are given by $\Psi_{n/(n+1)}=\mathcal{P}_{\rm LLL}\Phi_n\Phi_1$, where $\Phi_n$ is the wave function of $n$ filled LLs. In exact diagonalization studies for bosons interacting with a contact interaction (i.e., only the Haldane pseudopotential $V_0$ is nonzero), these are seen to provide a reasonably accurate description of the actual ground states at $\nu_b=1/2, 2/3, 3/4$, although the CF description becomes less accurate with increasing $n$ \cite{Cooper99,Regnault03, Chang05b}. If the CFs were noninteracting, one would expect a CF Fermi sea in the limit $n\rightarrow \infty$, i.e., at $\nu_b=1, 1/3$.  However, the bosonic ground state for the hard-core $V_0$  interaction at $\nu_b=1$ is not a CFFS but an incompressible state that has a high overlap with the $p$-wave paired MR-Pf state \cite{Regnault03, Chang05b}. 

We ask if our BCS formalism can capture this physics. In the spherical geometry, starting from the fermionic BCS state at $\nu=1/2$, a bosonic BCS state at $\nu_b=1$ can be written as \cite{Xie90}:
$
\Psi_{1}^{\rm bosonic-BCS} = \Psi^{\rm CF-BCS}_{1/2}/\Phi_{1}
$, 
where $\Phi_{1}$ is the wave function for one filled LL of electrons and $\Psi_{1/2}^{\rm CF-BCS}$ is the fermionic BCS state at $\nu=1/2$. In the torus geometry, the JK projected bosonic BCS state at $\nu_b=1$ is written as 
\begin{widetext}
\begin{equation}
    \Psi_{\nu_b=1}^{\rm bosonic-BCS} = e^{\sum_i \frac{z_i^2 - |z_i|^2}{4\ell^2}}\Bigg\{\vartheta
\begin{bmatrix}
{\phi_1\over 2\pi}
 + {N-1 \over 2}\\
-{\phi_2\over 2\pi } + {N-1 \over 2}
\end{bmatrix}
\Bigg({Z \over L} \Bigg | \tau \Bigg) \Bigg \} \frac{ {\rm Pf}(\tilde{M}_{ij})} {\prod_{\substack{ i<j}}  \vartheta  \begin{bmatrix}
{1 \over 2} \\ {1 \over 2 }
\end{bmatrix}\Bigg(\frac{z_i - z_j}{L}\Bigg |\tau \Bigg) },
\end{equation}
\end{widetext}
where $\tilde{M}_{ij}$ is given in Eq.~\eqref{JK2_Pf}. The above wave function has the same variational parameters as  Eq.~\eqref{BCS_paired_2}. The center of mass part is constructed at filling fraction $\nu_b=1$. The above wave function satisfies the proper quasi-periodic boundary conditions. 

We calculate the energy of the bosonic BCS state for two interactions: Coulomb, and $V_0$. The real space form of the $V_0$ interaction is given by $4\pi \delta^2(r)$ \cite{Regnault06}, for which we use the approximation
\beq
\delta^2(r) \equiv \lim _{\sigma \rightarrow 0} \frac{1}{2 \pi \sigma^2 }e^{-\frac{r^2}{2 \sigma^2}}
\eeq 
where $\sigma$ is the width of the Gaussian. We calculate the energy for multiple values of $\sigma$ and do an extrapolation to find the energy in the $\sigma \rightarrow 0$ limit. We calculate the energy per particle with respect to the variational parameters for a system of 12 bosons and minimize the energy by varying the two variational parameters: $\tilde{\Delta}$ and $k_{\rm cutoff}$. We find that in our calculations, the lowest energy state is obtained for $l=1$ pairing with $\tilde{\Delta} \neq 0$ and $k_{\rm cutoff} \neq k_F$. As shown in Fig.~\ref{fig:Bosonic-one}, the lowest energies (measured relative to the CFFS energy) are $\sim -0.02$ and $-0.009$ for the Coulomb and the $V_0$ interactions, respectively. This indicates the possibility of $p$-wave $(l=1)$ pairing for both Coulomb and the $V_0$ interaction, consistent with previous studies \cite{Regnault03, Chang05b}. For the $f$-wave $(l=3)$ CF-BCS state, the lowest energy state is obtained for $k_{\rm cutoff}=k_F$, which is the CFFS. 

In addition to the above state, an anti-Pfaffian bosonic state can be constructed by dividing the fermionic anti-Pfaffian wave function by the Jastrow factor \cite{Bose23}. This naturally raises the question if the anti-Pfaffian phase is competitive at $\nu_{b}=1$.  Because the anti-Pfaffian phase belongs to the $l=-3$ pairing, we look for pairing instability of the bosonic BCS state in the $l=-3$ channel. While it follows a similar trend as the BCS $l=1$ state, as shown in Fig.~\ref{fig:Bosonic-one}, the lowest energy is obtained for BCS $l=1$ state.

We have also studied this issue through exact diagonalization of the LLL Coulomb and the $V_{0}$ Hamiltonians for bosons at $\nu_{b}{=}1$ in the spherical geometry~\cite{Haldane83}. We find that the ground state at only the MR-Pf shift is consistently uniform i.e., has $L{=}0$ on the sphere, indicating that the thermodynamic ground state lies in the MR-Pf phase (see Table~\ref{tab: overlaps_exact_MR_nub_1}). In contrast, the ground state at the anti-Pfaffian shift is uniform for some particle numbers but not for others. We have not performed a systematic calculation of instability in the torus geometry as a function  of $N$.

{\it $\nu_b=\frac{1}{3}$:} The bosonic BCS state at $\nu_b=1/3$ can be written as $\Psi_{1/2}^{\rm CF-BCS}\Phi_1$. Following the modified JK projection scheme, we obtain the bosonic BCS wave function:
\begin{widetext}
\begin{equation}
    \Psi_{\nu_b=\frac{1}{3}}^{\rm bosonic-BCS} = e^{\sum_i \frac{z_i^2 - |z_i|^2}{4\ell^2}}\Bigg\{\vartheta
\begin{bmatrix}
{\phi_1\over 6\pi}
 + {N-1 \over 2}\\
-{\phi_2\over 2\pi } + {3(N-1) \over 2}
\end{bmatrix}
\Bigg({3Z \over L} \Bigg | 3\tau \Bigg) \Bigg \}  {\rm Pf}(\tilde{M}_{ij}) \prod_{\substack{ i<j}}  \vartheta  \begin{bmatrix}
{1 \over 2} \\ {1 \over 2 }
\end{bmatrix}\Bigg(\frac{z_i - z_j}{L} \Bigg|\tau \Bigg) 
\end{equation}
\end{widetext}
where $\tilde{M}_{ij}$ is given in Eq.~\eqref{JK2_Pf}. We calculate the energies for the BCS $l=1$, BCS $l=-3$ and BCS $l=3$ for the Coulomb interaction at $\nu_b=1/3$ as a function of $k_{\rm cutoff}$ and $\tilde{\Delta}$ and obtain the minimum energy states. We find the minimum energy state is obtained when $k_{\rm cutoff}=k_F$ for each value of $\tilde{\Delta}$ for BCS $l=1$, BCS $l=-3$ and BCS $l=3$. In other words, there is no pairing instability here. This is not surprising since the bosonic CFFS provides an excellent representation of the exact Coulomb ground state obtained from exact diagonalization studies \cite{Liu20}. Our findings suggest that the BCS formalism can also be used to explore bosonic states.

\begin{figure}[t]
		\includegraphics[width=\linewidth]{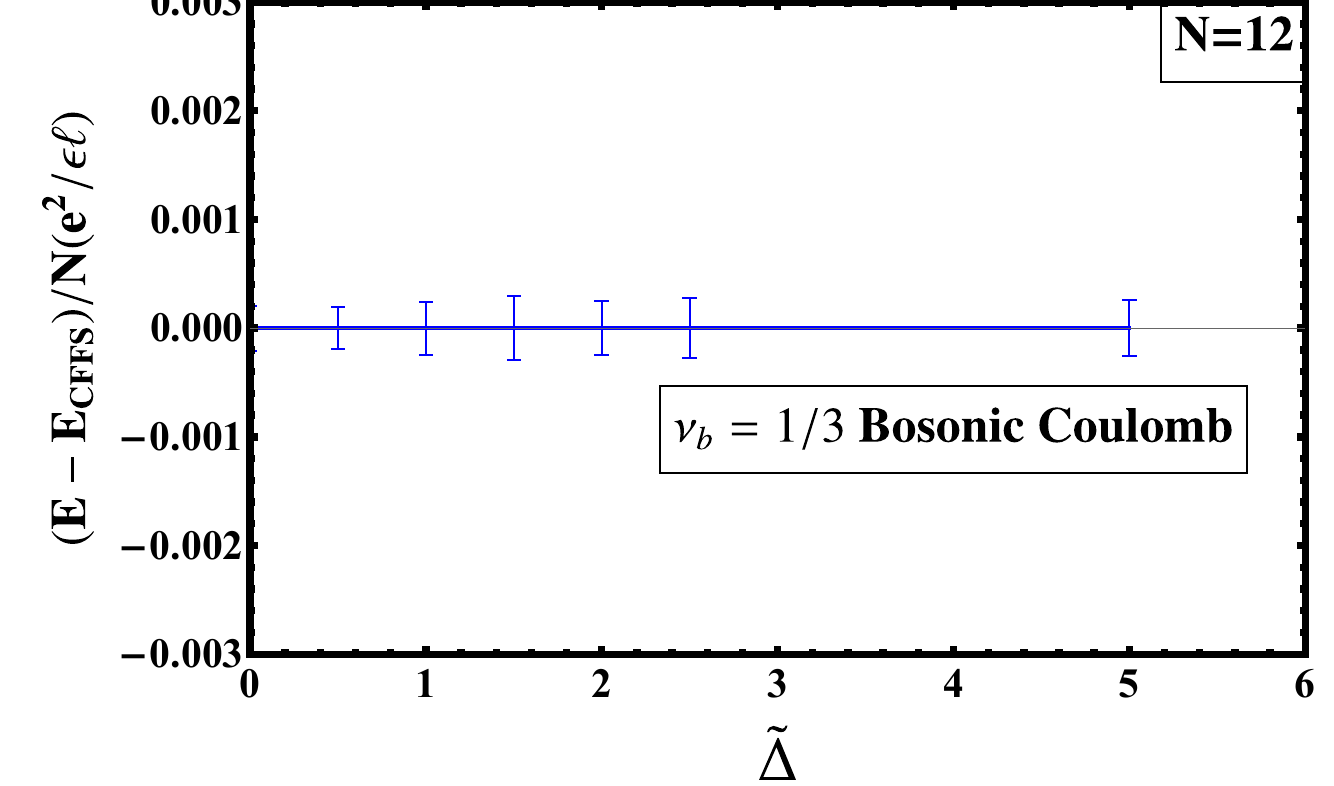}
		
\caption{\label{fig:Bosonic-one-third} The energy per particle for the bosonic BCS state with $l=1$ pairing at $\nu_b=1/3$ for a system of 12 bosons on a torus. The energies are plotted relative to the energy of the bosonic CFFS. The minimum energy state is obtained for $k_{\rm cutoff}=k_F$ for all values of $\tilde{\Delta}$, indicating the absence of a pairing instability. The same result is obtained for BCS $l=-3$ and BCS $l=3$ states.
}
\end{figure}

\section{Gap estimates from condensation energies}
\label{sec: gaps_from_Ec}

The variational parameter $\Delta$ ought not to be identified with the physical gap, $E_g$, of the paired CF state. 
However, it is tempting to ask if the condensation energy, i.e., the energy gain due to pairing of CFs, can provide an estimate for $E_g$ of the paired state.  According to the BCS theory, the condensation energy is given by~\cite{Schrieffer99}
\beq
E_{\rm CFFS}-E_{\rm CF-BCS}=N E_{\rm c}=\rho(E_{\rm F}) E_g^2/2,
\eeq
where $\rho(E_{\rm F})=Ak_F^2/(4\pi E_F)$ is the density of states at the Fermi energy ($A$ is the area) and $E_{\rm c}$ is the energy gain per particle due to pairing.
Using the relations $k_F=\sqrt{2\nu}/\ell$ and $N/A=\nu/(2\pi \ell^2)$, we get
\beq
E_g=\sqrt{2 E_c E_F}.
\label{eq:condensation}
\eeq
This is only a crude estimate of the physical gap, as we are assuming that the CFs can be modeled as weakly interacting fermions with a well defined mass with a quadratic dispersion.

At $\nu=1/2$, if we take the Fermi energy of the CFs to be 0.1 $e^2/\epsilon \ell$ (which is the estimated value at zero width~\cite{Mandal01a, Balram16b}), then a condensation energy of 0.002 $e^2/\epsilon \ell$ per particle yields $E_g\approx$ 0.02 $e^2/\epsilon \ell$. With finite-width the gap is expected to go down but currently we do not have a precise estimate of $E_{F}$ as a function of width. Since the charge gap at 1/3 at zero-width is also 0.1 $e^2/\epsilon \ell$~\cite{Fano86, Balram20b, Zhao22}, as a first approximation, one could use the 1/3 gap to estimate the Fermi energy of CFs. The 1/3 charge gap, as a function of the width and density, for a system of $N=12$ electrons obtained from exact diagonalization is given in Fig. D1(g) of Ref.~\cite{Zhao21}. Using that as an estimate of $E_{F}$, a condensation energy of 0.002 $e^2/\epsilon \ell$ per particle yields $E_g\lesssim$ 0.02 $e^2/\epsilon \ell$ for the widths and densities shown in Fig.~\ref{Phase diagram half}. 
	
Similarly, for $1/4$, we can approximate the $E_{F}$ as the charge gap at $1/5$. The charge gap at $1/5$ at zero-width is 0.02 $e^2/\epsilon \ell$~\cite{Balram21}. Using that as an estimate of $E_{F}$ (Note that we do not have an estimate of the 1/5 gap as a function of width and density.), a condensation energy of 0.01 $e^2/\epsilon \ell$ per particle at $w{=}70$ nm for $\rho{=}2\times 10^{11}$ cm$^{-2}$ (see Fig.~\ref{energy}) yields $E_g\approx$ 0.02 $e^2/\epsilon \ell$ for these parameters. 

\section{Conclusions}
\label{sec:conclusions}

In this article, we have applied the CF-BCS formalism to study phase transitions at filling factors $\nu=1/2$ and $\nu=1/4$ as a function of the QW width. We include the effect of finite width through an effective interaction between electrons, and find a $p$-wave instability at $\nu=1/2$ and an $f$-wave instability at $\nu=1/4$ as either the QW width or the density is increased. The phase diagram in the electron density - QW width plane is in excellent agreement with experiments at $\nu=1/2$ and in reasonable agreement at $\nu=1/4$. We note that the CFFS state is an excellent variational state at $\nu=1/2$ for narrow QWs, and therefore, the fact that we explicitly find a lower energy state for large QW widths is significant, and the agreement with experiments attests to the quantitative accuracy of our approach.

As noted above, the FQHE on either side of the $1/2$ FQHE is single-component like, showing the standard Jain fractions $n/(2n\pm 1)$, and the measured Fermi wave vector of CFs in close proximity to $\nu=1/2$ is also consistent with a single component CFFS~\cite{Mueed15}. Recent experiments have seen anomalously strong FQHE states at $8/17$ and $7/13$, which are 
consistent with the Levin-Halperin daughter states of the single component Pfaffian state at $\nu=1/2$~\cite{Levin09a}. We have shown in our present work that the single-component origin is also consistent with the disappearance of the 1/2 FQHE in QWs with a sufficient degree of asymmetry in the charge distribution. Calculations also show that the single-component paired CF state is a better variational state than the two-component Halperin 331 state in these wide QWs. These facts combined with the excellent agreement between our calculated and the measured transition boundaries separating the CFFS and the FQHE states strongly point to a one-component CF-pairing origin for the 1/2 FQHE in wide QWs. At $\nu=1/4$ as well, two-component FQHE states do not appear to be competitive~\cite{Faugno19}. 

Different pairing channels can in principle be distinguished through thermal Hall conductance, which is given by $\kappa = c \frac{\pi ^2 k_B^2}{3h} T$, where the chiral central charge is related to the relative angular momentum of the pair $l$ as $c=1+l/2$. Another quantity that can discriminate between the various candidate states is the Hall viscosity $\eta^A$ \cite{Avron95}, which is given by \cite{Read09} $\eta^A=\sh{\frac{\hbar} { 4}}{\rho}$, where $\rho$ is the 2D density and $\sh=N/\nu-N_\phi$ is the ``shift"  \cite{Wen92} in the spherical geometry.  The quantized values of $\sh$ for different candidate states, given by the relation $\sh=(2p+l)$, are listed in Table.~\ref{table-exp}. Another quantity of interest is the entanglement spectrum~\cite{Li08}, which contains information regarding the universal topological features of the phase. We note that Yutushui and Mross~\cite{Yutushui20} have considered CF-BCS wave function in the $l=-3$ channel in the spherical geometry and showed that its entanglement spectrum is consistent with that of the anti-Pf state. 

We have also applied the CF-BCS formalism to the problem of bosons in the lowest LL. At $\nu_b=1$ we find that the CFFS is unstable to $p$-wave pairing for both the contact and the Coulomb interactions, consistent with previous exact-diagonalization studies that support the MR-Pfaffian state. No such instability is found at $\nu_b=1/3$ for the Coulomb interaction, again in agreement with exact diagonalization studies.

We believe that the success of the CF-BCS theory in explaining a variety of experimental results makes a strong case for the CF pairing as the primary mechanism of FQHE at even-denominator fractions.

\section{Acknowledgments}

We are grateful to Mansour Shayegan for many useful discussions. A.S. and J.K.J acknowledge financial support from the U.S. National Science Foundation under grant no. DMR-2037990.  We acknowledge Advanced CyberInfrastructure computational resources provided by The Institute for CyberScience at The Pennsylvania State University and the Nandadevi supercomputer, which is maintained and supported by the Institute of Mathematical Science's High-Performance Computing Center. We thank M. Wimmer for the open-source PFAPACK library, used for the numerical evaluation of the Pfaffian of matrices. Some of the numerical calculations were performed using the AQUILLA~\cite{Martin20} and DiagHam~\cite{DiagHam} packages, for which we are grateful to its authors.

\begin{appendix}

\section{Periodic boundary conditions}
\label{pbc}
In this section, we show that the JK projected CF-BCS wave function given by Eq.~\eqref{BCS_paired_2} satisfies periodic boundary conditions given in Eq.~\eqref{eq:pbc}. For specificity, we shall consider the BCS wave function of $2N$ particles with positions $z_1,z_2,...,z_{2N}$ at $\nu=1/4$, i.e. subjected to $N_{\phi}=8N$ magnetic flux quanta. (Please bear in mind that the number of particles is denoted by $2N$ in this section.) It is straightforward to see that the boundary conditions in the $L_1$ direction are satisfied. Therefore, we consider here the boundary conditions in the $L_2$ direction. 

We consider the application of ordinary translation operator $T_p(L_2)$ on different parts of the wave function. Let us first consider $T_p(L_2)$ acting on $\tilde{M}_{ij}$, where we have two possibilities: (i) $i,j\neq p$ and (ii) $i$ or 
$j=p$. For $p\neq i,j$, we have:
\begin{widetext}
\begin{eqnarray}
&&  T_p(L\tau) \tilde{M}_{ij} \\\nonumber
&=&  
    \Bigg \{ \sum_{k_n}g_{k}e^{-\frac{\ell^2}{2}k_n(k_n+2\Bar{k}_n)}e^{\frac{i}{2}(z_i-z_j)(k_n+\Bar{k}_n)} e^{i\pi (\frac{2(z_i +i4k_n\ell^2 - z_p)}{L}-\tau)} e^{i\pi (\frac{2(z_j -i4k_n\ell^2 - z_p)}{L}-\tau)} \prod_{\substack{r \\r \neq i,j}} \left( \vartheta \begin{bmatrix} {1 \over 2} \\ {1 \over 2} \end{bmatrix}\Bigg(\frac{z_i + i4k_n\ell^2- z_r}{L}|\tau \Bigg)\right)\\\nonumber
   && \prod_{\substack{m \\m \neq i,j}} \left( \vartheta \begin{bmatrix} {1 \over 2} \\ {1 \over 2} \end{bmatrix}\Bigg(\frac{z_j - i4k_n\ell^2- z_m}{L}|\tau \Bigg) \right)  \Bigg(\vartheta \begin{bmatrix} {1 \over 2} \\ {1 \over 2} \end{bmatrix}\Bigg(\frac{z_i +i4k_n\ell^2 - z_j}{L}|\tau \Bigg)\Bigg)^2  \Bigg \}\\ \nonumber
    &=& e^{i2\pi \frac{(z_i+z_j)}{L}}e^{-i\frac{4\pi z_p}{L}}e^{-i2\pi \tau} \tilde{M}_{ij}. 
\end{eqnarray}

For $p=i$ or $p=j$, we get
\begin{eqnarray}
&& T_p(L \tau) \tilde{M}_{pj}\\\nonumber
 &=& T_p(L\tau) \Bigg \{ \sum_{k_n}g_{k_n}e^{-\frac{\ell^2}{2}k_n(k_n+2\Bar{k}_n)}e^{\frac{i}{2}(z_p-z_j)(k_n+\Bar{k}_n)}  \prod_{\substack{r \\r \neq p,j}}  \left(\vartheta \begin{bmatrix} {1 \over 2} \\ {1 \over 2} \end{bmatrix}\Bigg(\frac{z_p + i4k_n\ell^2- z_r}{L}|\tau \Bigg)\right) \\ \nonumber && \prod_{\substack{m \\m \neq p,j}}  \left(\vartheta \begin{bmatrix} {1 \over 2} \\ {1 \over 2} \end{bmatrix}\Bigg(\frac{z_j - i4k_n\ell^2- z_m}{L}|\tau \Bigg) \right)
 \Bigg(\vartheta \begin{bmatrix} {1 \over 2} \\ {1 \over 2} \end{bmatrix}\Bigg(\frac{z_p +i4k_n\ell^2 - z_j}{L}|\tau \Bigg)\Bigg)^2  \Bigg \} \\\nonumber
    &=&  \sum_{k_n}g_{k_n}e^{-\frac{\ell^2}{2}k_n(k_n+2\Bar{k}_n)}e^{\frac{i}{2}(z_p-z_j)(k_n+\Bar{k}_n)} e^{\frac{i}{2}L\tau(k+\Bar{k})} e^{-i (N-2)\pi( \frac{2(z_p+i4k_n\ell^2)}{L}+ \tau)}e^{i2\pi\frac{\sum_a' z_a}{L}} e^{-i2\pi(\frac{2(z_p+i4k_n\ell^2-z_j)}{L}+ \tau)}\\\nonumber
    &&\prod_{\substack{r \\r \neq p,j}} \left( \vartheta \begin{bmatrix} {1 \over 2} \\ {1 \over 2} \end{bmatrix}\Bigg(\frac{z_p + i4k_n\ell^2- z_r}{L}|\tau \Bigg)\right) \prod_{\substack{m \\m \neq p,j}} \left(\vartheta \begin{bmatrix} {1 \over 2} \\ {1 \over 2} \end{bmatrix}\Bigg(\frac{z_j - i4k_n\ell^2- z_m}{L}|\tau \Bigg) \right)  \Bigg(\vartheta \begin{bmatrix} {1 \over 2} \\ {1 \over 2} \end{bmatrix}\Bigg(\frac{z_p +i4k_n\ell^2 - z_j}{L}|\tau \Bigg)\Bigg)^2 \\\nonumber
    &=& \sum_{k_n}g_{k_n}e^{-\frac{l^2}{2}k_n(k_n+2\Bar{k_n})}e^{\frac{i}{2}(z_p-z_j)(k_n+\Bar{k_n})} e^{\frac{i}{2}L\tau(k_n+\Bar{k_n})} e^{-i N\pi( \frac{2z_p}{L}+\frac{i4k_nl^2}{L}+ \tau)}e^{i2\pi\frac{\sum_a' z_a}{L}} e^{i2\pi(\frac{2z_j}{L})}\\\nonumber
     &&\prod_{\substack{r \\r \neq p,j}} \left( \vartheta \begin{bmatrix} {1 \over 2} \\ {1 \over 2} \end{bmatrix}\Bigg(\frac{z_p + i4k_n\ell^2- z_r}{L}|\tau \Bigg)\right) \prod_{\substack{m \\m \neq p,j}} \left(\vartheta \begin{bmatrix} {1 \over 2} \\ {1 \over 2} \end{bmatrix}\Bigg(\frac{z_j - i4k_n\ell^2- z_m}{L}|\tau \Bigg) \right)  \Bigg(\vartheta \begin{bmatrix} {1 \over 2} \\ {1 \over 2} \end{bmatrix}\Bigg(\frac{z_p +i4k_n\ell^2 - z_j}{L}|\tau \Bigg)\Bigg)^2 \\\nonumber
     &=& e^{-i \frac{2N\pi z_p}{L}} e^{-iN\pi \tau} e^{i2\pi\frac{\sum_a' z_a}{L}} e^{i2\pi(\frac{2z_j}{L})} \tilde{M}_{pj}
\end{eqnarray}
where $\sum _a '= \sum_{\substack{ \\a\neq p,j}} $.
\end{widetext}

Remembering that the subscript $p$ appears in only one factor on the right hand side of Eq.~\eqref{eq:Pf}, 
the action of $T_p(L\tau)$ on the Pf[$\tilde{M_{ij}}$] yields:
\begin{equation}
T_p(L\tau){\rm Pf}(\tilde{M}_{ij})= e^{i\frac{4\pi Z}{L}} e^{-i\frac{4\pi Nz_p}{L}}e^{-i2N\pi \tau}e^{i2\pi \tau} {\rm Pf}(\tilde{M}_{ij})
\end{equation}

The action of $T_p(L\tau)$ on the Jastrow factor is given by
\begin{eqnarray}
\label{JK2_rel}
&& T_p(L\tau) \prod_{i<j} \vartheta  \Bigg\{\begin{bmatrix}
{1 \over 2} \\ {1 \over 2 }
\end{bmatrix}\Bigg(\frac{z_i - z_j}{L}|\tau \Bigg) \Bigg\}^2 \\ \nonumber
&=& e^{-i\frac{4\pi z_p (N-1)}{L}}e^{i4\pi\frac{\sum_a'' z_a}{L}} e^{-i2\pi \tau (N-1)} \prod_{i<j} \vartheta  \Bigg\{\begin{bmatrix}
{1 \over 2} \\ {1 \over 2 }
\end{bmatrix}\Bigg(\frac{z_i - z_j}{L}|\tau \Bigg) \Bigg\}^2 \\ \nonumber
&=& e^{-i4\pi N\frac{ z_p }{L}}e^{i4\pi\frac{Z}{L}} e^{-i2\pi \tau (N-1)} \prod_{i<j} \vartheta  \Bigg\{\begin{bmatrix}
{1 \over 2} \\ {1 \over 2 }
\end{bmatrix}\Bigg(\frac{z_i - z_j}{L}|\tau \Bigg) \Bigg\}^2
\end{eqnarray}
where $\sum_a''= \sum_{a \neq p}$. 

The translation of the center of mass gives the relation:
\begin{widetext}
\begin{equation}
\label{com}
T_p(L\tau)\Bigg\{\vartheta
\begin{bmatrix}
{\phi_1\over 8\pi}
 + {N-1 \over 2}\\ 
-{\phi_2\over 2\pi } + {2(N-1)}
\end{bmatrix}
\Bigg({4Z \over L} \Bigg |4 \tau \Bigg) \Bigg \}  = e^{i\phi _{\tau}} e^{-i4\pi \tau} e^{-i\frac{8\pi Z}{L}}  \Bigg\{\vartheta
\begin{bmatrix}
{\phi_1\over 8\pi}
 + {N-1 \over 2}\\ 
-{\phi_2\over 2\pi } + {2(N-1)}
\end{bmatrix}
\Bigg({4Z \over L} \Bigg |4 \tau \Bigg) \Bigg \} 
\end{equation}

Combining all the factors obtained in Eq.~\eqref{JK2_Pf}, Eq.~\eqref{JK2_rel} and Eq.~\eqref{com}, we find that 
\begin{eqnarray}
 T_p(L\tau) \Bigg\{\vartheta
\begin{bmatrix}
{\phi_1\over 8\pi}
 + {N-1 \over 2}\\ 
-{\phi_2\over 2\pi } + {2(N-1)}
\end{bmatrix}
\Bigg({4Z \over L} \Bigg |4 \tau \Bigg) \Bigg \} {\rm Pf}(\tilde{M}_{ij})  \prod_{i<j} \vartheta  \Bigg\{\begin{bmatrix}
{1 \over 2} \\ {1 \over 2 }
\end{bmatrix}\Bigg(\frac{z_i - z_j}{L}|\tau \Bigg) \Bigg\}^2  \nonumber \\= e^{i\phi_2} e^{-i\pi N_{\phi}(\frac{2z_p}{L}+\tau)} \Bigg\{\vartheta
\begin{bmatrix}
{\phi_1\over 8\pi}
 + {N-1 \over 2}\\ 
-{\phi_2\over 2\pi } + {2(N-1)}
\end{bmatrix}
\Bigg({4Z \over L} \Bigg |4 \tau \Bigg) \Bigg \}  \nonumber
{\rm Pf}(\tilde{M}_{ij})  \prod_{i<j} \vartheta  \Bigg\{\begin{bmatrix}
{1 \over 2} \\ {1 \over 2 }
\end{bmatrix}\Bigg(\frac{z_i - z_j}{L}|\tau \Bigg) \Bigg\}^2
\end{eqnarray}
\end{widetext}
Using Eq.~\eqref{eq:L2}, we have for magnetic translation operator
\begin{equation}
 t_p(L\tau) \Psi^{\rm BCS}=e^{i\phi_2}\Psi^{\rm BCS}
\end{equation}
which is the desired quasi-periodic boundary condition along the $\tau$ direction. 

For completeness, we now show the JK projected wave fucntion obtained by bringing all the Jastrow factors inside the Pfaffian matrix also preserves the PBCs.
\begin{equation}
{\rm Pf}\left[\sum_ng_{\vec{k}_n}\hat{F}_n(z_i,z_j)\right]\prod_iJ_i^2\rightarrow 
{\rm Pf}\left[\sum_ng_{\vec{k}_n}\hat{F}_n(z_i,z_j)J_i^2 J_j^2\right]
\end{equation}
\begin{equation}
    \Psi_{\frac{1}{4}}^{\rm BCS} = e^{\sum_i \frac{z_i^2 - |z_i|^2}{4\ell^2}}\Bigg\{\vartheta
\begin{bmatrix}
{\phi_1\over 8\pi}
 + {N-1 \over 2}\\ 
-{\phi_2\over 2\pi } + {2(N-1)}
\end{bmatrix}
\Bigg({4Z \over L} \Bigg |4 \tau \Bigg) \Bigg \} {\rm Pf}(M^*_{ij})
\end{equation}
in which the Pfaffian matrix element is:

\begin{widetext}
\begin{eqnarray}
M^*_{ij}=\sum_{k_n}g_{\vec{k}_n}e^{-\frac{\ell^2}{2}k_n(k_n+2\Bar{k}_n)}e^{\frac{i}{2}(z_i-z_j)(k_n+\Bar{k}_n)} \prod_{\substack{m \\m \neq i,j}} \left(  \vartheta  \begin{bmatrix}
{1 \over 2} \\ {1 \over 2 }
\end{bmatrix}\Bigg(\frac{z_i + i\textcolor{red}{2}k_n\ell^2- z_m}{L}|\tau \Bigg)  \right)^2 \nonumber \\ \prod_{\substack{n \\n \neq i,j}} \Bigg( \vartheta  \begin{bmatrix}
{1 \over 2} \\ {1 \over 2 }
\end{bmatrix}\Bigg(\frac{z_j - i\textcolor{red}{2}k_n\ell^2- z_n}{L}|\tau \Bigg) \Bigg )^2  \Bigg \{\vartheta  \begin{bmatrix}
{1 \over 2} \\ {1 \over 2 }
\end{bmatrix}\Bigg(\frac{z_i + i\textcolor{red}{2}k_n\ell^2- z_j}{L}|\tau \Bigg) \Bigg \}^4 \nonumber
\end{eqnarray}
Here $M^*$ does not indicate complex conjugate of $M$. The translation of the CM part of the wave function gives us the relation:

\begin{equation}
T_p(L\tau)\Bigg\{\vartheta
\begin{bmatrix}
{\phi_1\over 8\pi}
 + {N-1 \over 2}\\ 
-{\phi_2\over 2\pi } + {2(N-1)}
\end{bmatrix}
\Bigg({4Z \over L} \Bigg |4 \tau \Bigg) \Bigg \}  = e^{i\phi _{\tau}} e^{-i4\pi \tau} e^{-i\frac{8\pi Z}{L}}  \Bigg\{\vartheta
\begin{bmatrix}
{\phi_1\over 8\pi}
 + {N-1 \over 2}\\ 
-{\phi_2\over 2\pi } + {2(N-1)}
\end{bmatrix}
\Bigg({4Z \over L} \Bigg |4 \tau \Bigg) \Bigg \} 
\end{equation}
The translation of a single matrix element along the $\tau$ direction gives us, for $p\neq i,j$:
\begin{eqnarray}
&&  T_p(L\tau) M^*_{ij} \\\nonumber
&=&  
    \Bigg \{ \sum_{k_n}g_{k}e^{-\frac{\ell^2}{2}k_n(k_n+2\Bar{k}_n)}e^{\frac{i}{2}(z_i-z_j)(k_n+\Bar{k}_n)} e^{i2\pi (\frac{2(z_i +i2k_n\ell^2 - z_p)}{L}-\tau)} e^{i2\pi (\frac{2(z_j -i2k_n\ell^2 - z_p)}{L}-\tau)} \prod_{\substack{r \\r \neq i,j}} \left( \vartheta \begin{bmatrix} {1 \over 2} \\ {1 \over 2} \end{bmatrix}\Bigg(\frac{z_i + i2k_n\ell^2- z_r}{L}|\tau \Bigg)\right)^2\\\nonumber
   && \prod_{\substack{m \\m \neq i,j}} \left( \vartheta \begin{bmatrix} {1 \over 2} \\ {1 \over 2} \end{bmatrix}\Bigg(\frac{z_j - i2k_n\ell^2- z_m}{L}|\tau \Bigg) \right)^2  \Bigg(\vartheta \begin{bmatrix} {1 \over 2} \\ {1 \over 2} \end{bmatrix}\Bigg(\frac{z_i +i2k_n\ell^2 - z_j}{L}|\tau \Bigg)\Bigg)^4  \Bigg \}\\ \nonumber
    &=& e^{i4\pi \frac{(z_i+z_j)}{L}}e^{-i\frac{8\pi z_p}{L}}e^{-i4\pi \tau} M^*_{ij} 
\end{eqnarray}
and for $p=i$ or $p=j$, we get
\begin{eqnarray}
&& T_p(L \tau) M^*_{pj}\\\nonumber
 &=& T_p(L\tau) \Bigg \{ \sum_{k_n}g_{k_n}e^{-\frac{\ell^2}{2}k_n(k_n+2\Bar{k}_n)}e^{\frac{i}{2}(z_p-z_j)(k_n+\Bar{k}_n)}  \prod_{\substack{r \\r \neq p,j}}  \left(\vartheta \begin{bmatrix} {1 \over 2} \\ {1 \over 2} \end{bmatrix}\Bigg(\frac{z_p + i2k_n\ell^2- z_r}{L}|\tau \Bigg)\right)^2 \\ \nonumber 
 && \prod_{\substack{m \\m \neq p,j}}  \left(\vartheta \begin{bmatrix} {1 \over 2} \\ {1 \over 2} \end{bmatrix}\Bigg(\frac{z_j - i2k_n\ell^2- z_m}{L}|\tau \Bigg) \right)^2
 \Bigg(\vartheta \begin{bmatrix} {1 \over 2} \\ {1 \over 2} \end{bmatrix}\Bigg(\frac{z_p +i2k_n\ell^2 - z_j}{L}|\tau \Bigg)\Bigg)^4  \Bigg \} \\\nonumber
&=& e^{-i \frac{4N\pi z_p}{L}} e^{-i2N\pi \tau} e^{i4\pi\frac{\sum_a' z_a}{L}} e^{i4\pi(\frac{2z_j}{L})} M^*_{pj}
\end{eqnarray}
where $\sum _a '= \sum_{\substack{ \\a\neq p,j}} $.

\begin{equation}
    T_p(L\tau)\Bigg\{\vartheta
\begin{bmatrix}
{\phi_1\over 8\pi}
 + {N-1 \over 2}\\ 
-{\phi_2\over 2\pi } + {2(N-1)}
\end{bmatrix}
\Bigg({4Z \over L} \Bigg |4 \tau \Bigg) \Bigg \} {\rm Pf}(M^*_{ij}) = e^{i(\phi _{\tau}-N_{\phi}\pi (\frac{2z_p}{L} + \tau))}\Bigg\{\vartheta
\begin{bmatrix}
{\phi_1\over 8\pi}
 + {N-1 \over 2}\\ 
-{\phi_2\over 2\pi } + {2(N-1)}
\end{bmatrix}
\Bigg({4Z \over L} \Bigg |4 \tau \Bigg) \Bigg \} {\rm Pf}(M^*_{ij})
\end{equation}
which is exactly what the periodic boundary condition requires. In the other direction, the periodic boundary condition is satisfied in a similar way.
\end{widetext}
\section{Momentum sector for the wave functions}
\label{momentum-sector}
In this section, we try to find the momentum sector for the MR type states, labeled by their Haldane pseudomomemntum $(K_x,K_y)$. The Haldane pseudomomentum are given by the eigenvalues of the relative translation operator \cite{Haldane85b, Bernevig12}, which is given by 
\begin{equation}
\tilde{t}_i(L_j/N) = \prod_{k=1}^N t_i(L_j/N) t_k(-L_j/N).
\end{equation}
Here, we try to find the momentum sector for the MR-Pf state at $\nu=1/4$. The center of mass part of the wave function in Eq.~\eqref{MR} is invariant under the action of the relative translation operator. If we consider the action of the relative translation operator on the Pfaffian part of the MR-Pf wave function, we obtain the relation:
\begin{widetext}
\begin{equation}
\tilde{t}_i(L/N){\rm Pf}\Bigg(\frac{\vartheta \begin{bmatrix}
{a} \\ {b}
\end{bmatrix}
\Bigg( {z_i-z_j\over L}\Bigg | \tau \Bigg )}{\vartheta 
\begin{bmatrix}
{1\over 2} \\ {1\over 2}
\end{bmatrix}
\Bigg( {z_i-z_j\over L}\Bigg | \tau \Bigg )}\Bigg) = e^{i 2 \pi (a-\frac{1}{2})}{\rm Pf}\Bigg(\frac{\vartheta \begin{bmatrix}
{a} \\ {b}
\end{bmatrix}
\Bigg( {z_i-z_j\over L}\Bigg | \tau \Bigg )}{\vartheta 
\begin{bmatrix}
{1\over 2} \\ {1\over 2}
\end{bmatrix}
\Bigg( {z_i-z_j\over L}\Bigg | \tau \Bigg )}\Bigg)
\end{equation}.
\end{widetext}
The action of the relative translation operator on a single Jastrow factor: $\tilde{t}_i(L/N) \vartheta 
\begin{bmatrix}
{1\over 2} \\ {1\over 2}
\end{bmatrix}
\Bigg( {z_i-z_j\over L}\Bigg | \tau \Bigg ) \rightarrow  \vartheta 
\begin{bmatrix}
{1\over 2} \\ {1\over 2}
\end{bmatrix}
\Bigg( {z_i-z_j\over L}+1\Bigg | \tau \Bigg )  $. The action of the operator $\tilde{t}_i(L/N)$ on terms of the form $ \vartheta 
\begin{bmatrix}
{1\over 2} \\ {1\over 2}
\end{bmatrix}
\Bigg( {z_p-z_q\over L}\Bigg | \tau \Bigg )$ where $p,q \neq i$ leaves the term invariant.
Thus, the action of $\tilde{t}_i(L/N)$ on the Jastrow factors gives a factor of 1. Putting all the terms together, we obtain
\begin{equation}
\tilde{t}_i(L/N) \Psi^{(a,b)}_{{\rm MR}-p}=e^{i2\pi (a-\frac{1}{2})} \Psi^{(a,b)}_{{\rm MR}-p} = e^{-i2\pi\frac{K_x }{N}} \Psi^{(a,b)}_{{\rm MR}-p}
\end{equation}
In the other direction, we obtain the relation:
\begin{equation}
\tilde{t}_i(L\tau/N) \Psi^{(a,b)}_{{\rm MR}-p}=e^{-i2\pi (b-\frac{1}{2})} \Psi^{(a,b)}_{{\rm MR}-p}= e^{-i2\pi\frac{K_y }{N}} \Psi^{(a,b)}_{{\rm MR}-p}
\end{equation}
$(a,b)$ can take values $(0,\frac{1}{2}),(\frac{1}{2},0)$ or $(0,0)$, which correspond to $(K_x,K_y)=(N/2,0),(0,N/2)$, and $(N/2,N/2)$.

\section{Testing paired states in the spherical geometry}
\label{appx-sphere}

In the course of this work, we have also performed calculations in Haldane's spherical geometry\cite{Haldane83}, which we discuss in this Appendix. In this geometry, different candidate states at a given filling factor occur, in general, at different $N$ and different flux values, and therefore the electron-electron interaction energies of finite systems may not be directly compared. To obtain the thermodynamic limits of the energies one must include the contribution coming from the positively charged neutralizing background. Two previous works that dealt with finite width effects~\cite{Faugno19,Zhao21} had assumed that the electron-background and background-background energies can be evaluated by assuming that the electron as well as the neutralizing background charge was purely two-dimensional. The expectation was that the nature of the neutralizing background should not affect the energy {\it differences} between the various candidate states at a given filling factor. This would be true if sufficiently large systems could be considered. However, as the following discussion shows, that is not the case for finite systems and the model used may affect the thermodynamic limit. In this Appendix we consider a model in which the background charge has the same finite width distribution as the electron charge. This yields much better linear fits for the energy as a function of $1/N$ and hence, we believe, produces more reliable thermodynamic values. With this method, the phase boundary obtained in Ref.~\cite{Faugno19} at $\nu=1/4$ is slightly modified, although a transition into the $f$-wave 22111 still occurs. On the other hand, in contrast to the claim in Ref.~\cite{Zhao21}, the Pfaffian state has higher energy than the CFFS at $\nu=1/2$ in the entire range of width and density studied.

We note that this is not an issue for the torus geometry, because here all states at a given filling factor occur at the same flux, and therefore the electron-background and background-background terms exactly cancel for any finite system when energy differences are determined, independent of what model is used for the background charge.

We consider $N$ particles confined to the surface of a sphere subjected to $2Q$ magnetic flux quanta \cite{Haldane83} perpendicular to the surface. The radial magnetic field is originating from a magnetic monopole at the center of the sphere. The positions of the electrons can be denoted by spinor coordinates $u=cos(\theta)e^{i\phi/2}$ and $v=sin(\theta)e^{-i\phi/2}$, where $\theta$ and $\phi$ are the polar and azimuthal angles respectively. The radius of the sphere is $R=\sqrt{Q}\ell$. The distance of any two particles $i$ and $j$ is given by the chord length, which is equal to $\sqrt{Q}\ell|u_iv_j-u_jv_i|$.

For a FQH state at $\nu=n/(2pn+1)$, the composite fermions feel an effective magnetic field originating from a monopole  of strength $2Q^*=N/n-n$ where $n$ is the number of filled LLs. The relation between the total magnetic field and number of particles is as follows: $2Q = N/\nu - \mathcal{S}$, where $\mathcal{S}$ is called the ``shift." The filling factor in spherical geometry is defined as $\nu = {\rm lim}_{N\rightarrow \infty} N/2Q$.

The total energy includes a contribution from the positively charged uniform background. Previously~\cite{Faugno19, Zhao21}, the sum of the electron-background and the background-background contributions was taken to be $-N^2/2\sqrt{Q}$ in Coulomb units, which assumes the interaction to be of the bare Coulomb form. However, if we assume that the positively charged neutralizing background has the same charge distribution as the electrons as a function of the width, the background contribution for the finite-width interaction is different from that ascertained from the bare Coulomb form. For an arbitrary interaction $V(r)$ the contribution of the positively charged background (i.e. the sum of electron-background and background-background interactions) is given by~\cite{Balram20b, Bose23}
\begin{equation}
			E_{b} =-\frac{N^{2}}{4} \int_0^{\pi} \sin(\theta)d\theta~V \left( 2 R \sin \left(\frac{\theta}{2} \right) \right),   
			\label{eq: background_contribution}
\end{equation}
where $r=2R \sin(\theta/2)$ is the chord distance on the sphere with respect to which we evaluate energies. (The above equation is written for a single component system, but can be readily generalized to the two-component state.) The density-corrected per-particle total energy, which is what we extrapolate as a linear function of $1/N$ to the thermodynamic limit, is $E^{\rm pp}_{\rm tot}{=}\sqrt{2Q\nu/N} \left(E_{e-e}+E_{b}\right)/N$, where $E_{e-e}$ is the electron-electron contribution. We find that incorporating the background contribution using Eq.~\eqref{eq: background_contribution} significantly improves the linear fit of the $E^{\rm pp}_{\rm tot}$ as a function of $1/N$, indicating that it provides better thermodynamic limits. 

As before, we use transverse wave functions obtained using LDA and do not consider any LL mixing in the calculations reported in this section.

{\it $\nu=1/2$}: We have considered the competition between CFFS, MR state and the Halperin 331 states at $\nu=1/2$ in spherical geometry.
The MR wave function at $\nu=1/2$ in spherical geometry is given by:
\beq
\Psi_{\rm MR-{\it p}} = {\rm Pf}\left(\frac{1}{u_iv_j-u_jv_i}\right)\prod_{i<j} (u_iv_j-u_jv_i)^2.
\eeq
The Halperin 331 state is obtained by replacing $z_i-z_j$ by $(u_iv_j-u_jv_i)$ in Eq.~\eqref{eq:331}. As shown in Fig.~\ref{sphere-one-half-therm}, we find that in the thermodynamic limit, the CFFS has lower energy than the MR and 331 states for all values of densities and well widths considered for our numerical calculations. (In this case, the model used for the background charge leads to a qualitatively different conclusion than that in Ref.~\cite{Zhao21}.)

{\it $\nu=1/4$}: We consider the competition between the CFFS, MR-$p$ and the  $22111$ parton state which are given as \cite{Halperin93,Rezayi94,Faugno19}:
\beq
\Psi_{\rm CFFS} = \mathcal{P}_{\rm LLL} \Phi_{FS} \Phi_1^4
\eeq
\beq
\Psi_{\rm MR-{\it p}} = {\rm Pf}\left(\frac{1}{u_iv_j-u_jv_i}\right)\prod_{i<j} (u_iv_j-u_jv_i)^4
\eeq
\beq
\Psi_{\rm 22111} =\mathcal{P}_{\rm LLL} \Phi_2 \Phi_2 \Phi_1^3
\eeq

\begin{figure}[h]
		\includegraphics[width=\linewidth]{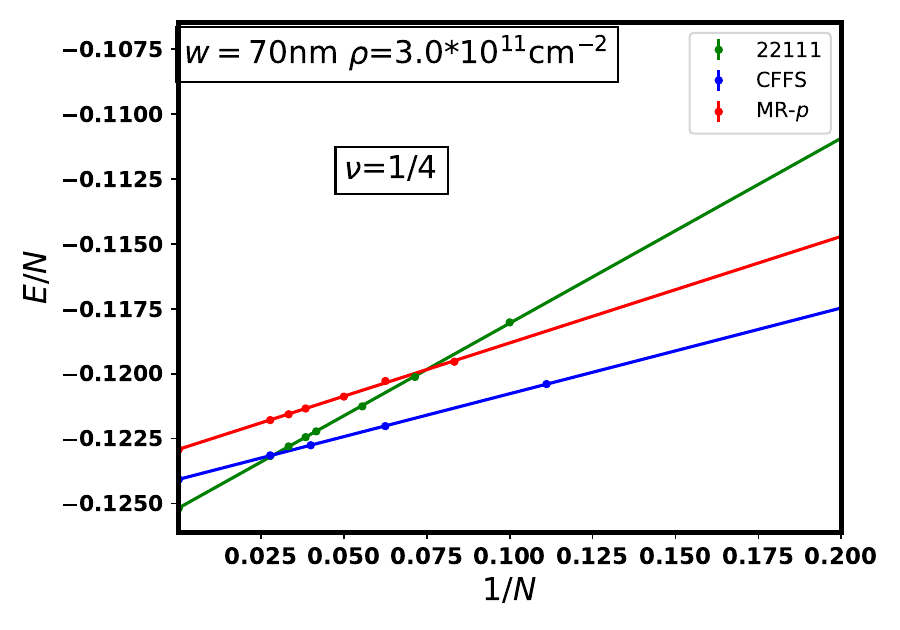}		
\caption{\label{sphere-one-fourth-therm} The energy $E/N$ for the CFFS and the 22111 states at $\nu=1/4$ as a function of $1/N$. The results are for a QW width $w=70$ nm and density $\rho=3.0 \times 10^{11}$ cm$^{-2}$. The spherical geometry is used for the calculation. The 22111 state has lower energy than the CFFS in the thermodynamic limit. The error bars are smaller than the size of the symbols. The energies are plotted in units of $e^2/\epsilon \ell$.
}
\end{figure}

\begin{figure}[h]
		\includegraphics[width=\linewidth]{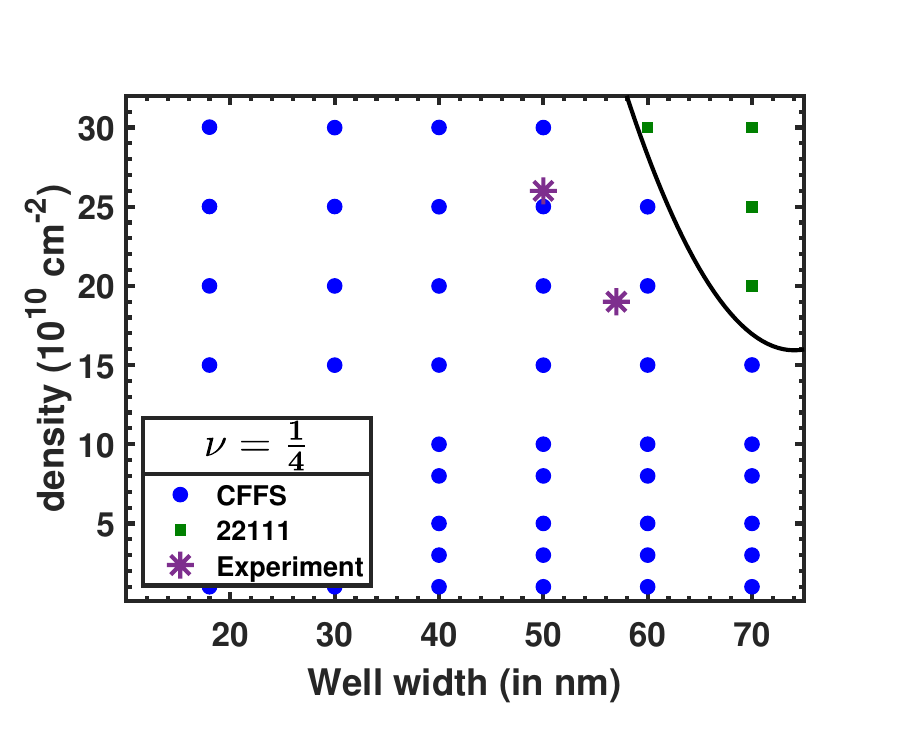}		
\caption{\label{sphere-one-fourth} The phase diagram indicating the lowest energy state at $\nu=1/4$ as a function of density and quantum-well width. The solid line represents the approximate theoretical phase boundary. The stars  indicate the parameter values where experiments show the onset of FQHE  at $\nu=1/4$.}
\end{figure}

Since the universality class of the wave function as well as its microscopic form is not very sensitive to the details of the projection~\cite{Balram16b, Anand22}, we project $\Psi_{\rm 22111}$ to the LLL as $\Psi_{2/5}^2/\Phi_1$, which allows for its evaluation for large system sizes using the JK projection of $\Psi_{2/5}$. The $22111$ state is an $f$-wave paired state with $l=3$~\cite{Balram18, Faugno19}. The shift for the 22111 state is ${\cal S}=7$. In Fig.~\ref{sphere-one-fourth}, we show the lowest energy state as a function of density and well width. For each data point in the phase diagram, we obtain the thermodynamic per particle energies by extrapolating the per perticle energies of finite systems as shown in Fig.~\ref{sphere-one-fourth-therm}. The energies are in units of $e^2/\epsilon \ell$. We find that the CFFS has lower energy for small well widths and low densities. However, for very large well width and densities, we find that the $22111$ parton state has lower energy. We 
find that the $MR$-p state always has higher energy than the CFFS and the $22111$ states in the thermodynamic limit, in agreement with the result reported in Ref.~\cite{Faugno19}. We note that while we still see a transition into the 22111 state as a function of increasing width or density, the phase boundary has shifted as compared to that in Ref.~\cite{Faugno19} because of the different models for background charge. 

{\it $\nu_{b}=1$ bosons}: For completeness, we have evaluated the overlaps of the bosonic MR state at $\nu_{b}{=}1$ in the spherical geometry with the exact ground state of the $\delta$-function and Coulomb interactions. For reference, the $\delta$-function interaction on the sphere corresponds to a pseudopotential of $V_{0}^{\delta-{\rm function}}{=}(1 {+}2Q)^2/\left[4 \pi Q \left(1 {+}4Q \right)\right]$ (with $V_{m}^{\delta-{\rm function}}{=}0~\forall~m{>}0$) which in the thermodynamic limit $Q{\to}\infty$ implies that the $V_{m}{=}\delta_{m,0}$ pseudopotential interaction (referred to from here on in as the $V_{0}$ interaction) corresponds to the interaction $V(\vec{r}){=}4\pi \delta(\vec{r})$ in real space. On the sphere, the MR wave function describing bosons at $\nu_{b}=1$ is given by:
\beq
\Psi^{\rm MR-{\it p}}_{\nu_{b}=1} = {\rm Pf}\left(\frac{1}{u_iv_j-u_jv_i} \right)\prod_{i<j} (u_iv_j-u_jv_i).
\label{eq: MR_Pf_nub_1}
\eeq
In Table~\ref{tab: overlaps_exact_MR_nub_1} we present the overlaps of the ground states of the $V_{0}$ and the LLL Coulomb Hamiltonians with the MR wave function for bosons at $\nu_{b}=1$. We find that the overlaps are reasonably high for all systems we have considered suggesting that the ground state of short-range dominated interactions for bosons at $\nu_{b}=1$ resides in the MR phase. Some of the numbers shown in Table~\ref{tab: overlaps_exact_MR_nub_1}, for systems smaller than the largest ones considered here, were previously given in Refs.~\cite{Regnault03, Chang05b}. 

\begin{table}
	\begin{center}	
	\begin{tabular}{|c|c|c|c|}
	\hline
	$N$	&	$|\langle \Psi^{V_{0}}_{1} | \Psi^{\rm MR-{\it p}}_{1} \rangle |^{2}$	&	$|\langle \Psi^{{\rm C(S)}}_{1} | \Psi^{\rm MR-{\it p}}_{1} \rangle |^{2}$	&	$|\langle \Psi^{{\rm C(D)}}_{1} | \Psi^{\rm MR-{\it p}}_{1} \rangle |^{2}$	\\ \hline
	4	&	1.0000	&	1.0000	&	1.0000	\\ \hline
	6	&	0.9728	&	0.9728	&	0.9728	\\ \hline
	8	&	0.9669	&	0.9771	&	0.9625	\\ \hline
	10	&	0.9592	&	0.9659	&	0.9618	\\ \hline
	12	&	0.8844	&	0.9165	&	0.9230	\\ \hline
	14	&	0.8858	&	0.9213	&	0.9156	\\ \hline
	16	&	0.8833	&	0.9170	&	0.9127	\\ \hline
	18	&	0.8504	&	0.8926	&	0.8923	\\ \hline
	20	&	0.7885	&	0.8599	&	0.8621	\\ \hline
\end{tabular}
\end{center}
	\caption{Squared overlaps of the exact $V_{0}$ and LLL Coulomb (C) ground states [obtained using the spherical (S) and planar disk (D) pseudopotentials] with the $\nu_{b}={1}$ MR Pf wave function of Eq.~\eqref{eq: MR_Pf_nub_1} for $N$ bosons in the spherical geometry.  This table includes results from Refs.~\cite{Regnault03, Chang05b} for completeness.}
	\label{tab: overlaps_exact_MR_nub_1}
\end{table}

Next, to check the incompressibility of the $\nu_{b}{=}1$ state, we compute its neutral and charge gaps for the $V_{0}$ and Coulomb interactions using exact diagonalization in the spherical geometry. The neutral gap $\Delta^{\rm neutral}$ is defined as the difference between the two lowest energies for a given system of $N$ electrons at the MR flux of $2Q{=}N{-}2$. The charge gap is defined as~\cite{Balram20b}
\begin{eqnarray}
	\Delta^{\rm charge}&=&\frac{\mathcal{E}(2Q-1)+\mathcal{E}(2Q+1)-2\mathcal{E}(2Q)}{n_{q}}, \\
	\mathcal{E}(2Q-1)&=&E(2Q-1)-(N^{2}-(n_{q} e_{q})^2) \frac{\mathcal{C}(2Q-1)}{2}, \nonumber \\ 
	\mathcal{E}(2Q   )&=&E(2Q)-(N^{2}           ) \frac{\mathcal{C}(2Q)}{2},  \nonumber \\ 
	\mathcal{E}(2Q+1)&=&E(2Q+1)-(N^{2}-(n_{q}e_{q})^2)\frac{\mathcal{C}(2Q+1)}{2},  \nonumber
	\label{eq:charge_gap}
\end{eqnarray}
where $\mathcal{C}(2Q)$ is the charging energy that accounts for the background, $E(2Q)$ is electron-electron interaction energy of the ground state obtained from exact diagonalization of $N$ electrons at flux $2Q$, $n_{q}{=}2$ is the number of quasiholes produced when a single flux quantum is inserted in the MR state and $e_{q}{=}1/2$ is the charge of the fundamental MR quasihole in units of the electronic charge. The $N^2$ term accounts for the background contribution while the $(n_{q} e_{q})^2$ term corrects for the fact that in the presence of additional charge in the form of quasiholes or quasiparticles the background is different~\cite{Jain07}. The charging energies $\mathcal{C}(2Q)$ of various interactions considered here at flux $2Q$ are given by~\cite{Balram20b}
\begin{eqnarray}
	\mathcal{C}\left(V_{m}{=}\delta_{m,0}\right) &=& \frac{(4Q+1)}{(2Q+1)^{2}}~\frac{e^{2}}{\epsilon \ell} \\
	\mathcal{C}^{\rm sphere}\left(\frac{1}{r}\right) &=& \frac{1}{\sqrt{Q}}~\frac{e^{2}}{\epsilon \ell} \nonumber  \\
	\mathcal{C}^{\rm disk}\left(\frac{1}{r}\right) &=& \frac{(3+4(2Q))\Gamma[2Q+3/2]}{3(2Q+1)^{2}\Gamma[2Q+1]}~\frac{e^{2}}{\epsilon \ell}, \nonumber 
	\label{eq:charging_energies}
\end{eqnarray}
where $\Gamma[x]$ is the Gamma function. The gaps are density-corrected~\cite{Morf86} by a factor of $\sqrt{2Q\nu/N}$ before extrapolation to the thermodynamic limit. 

\begin{figure}[htpb!]
	\begin{center}
		\includegraphics[width=0.47\textwidth,height=0.23\textwidth]{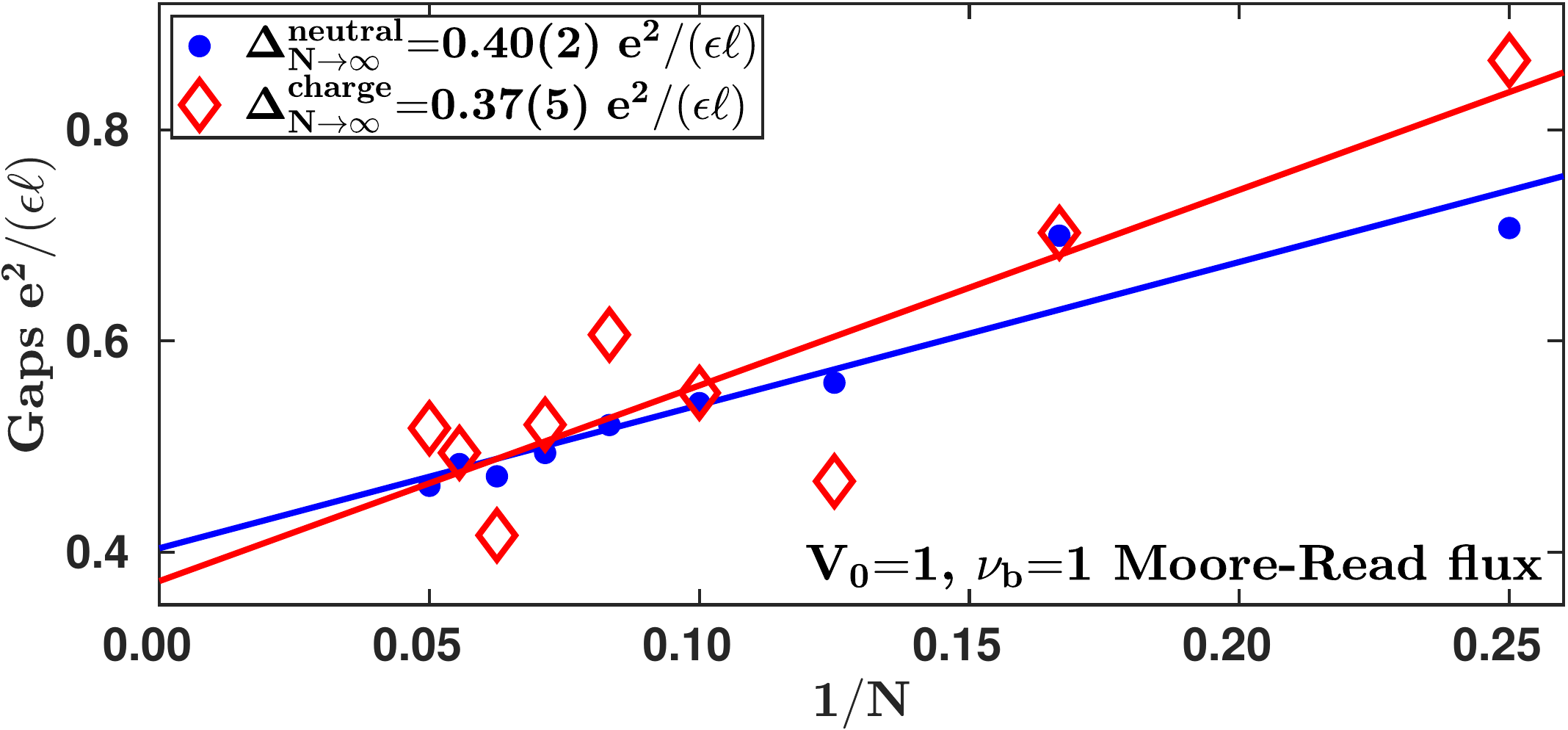} \\
		\includegraphics[width=0.47\textwidth,height=0.23\textwidth]{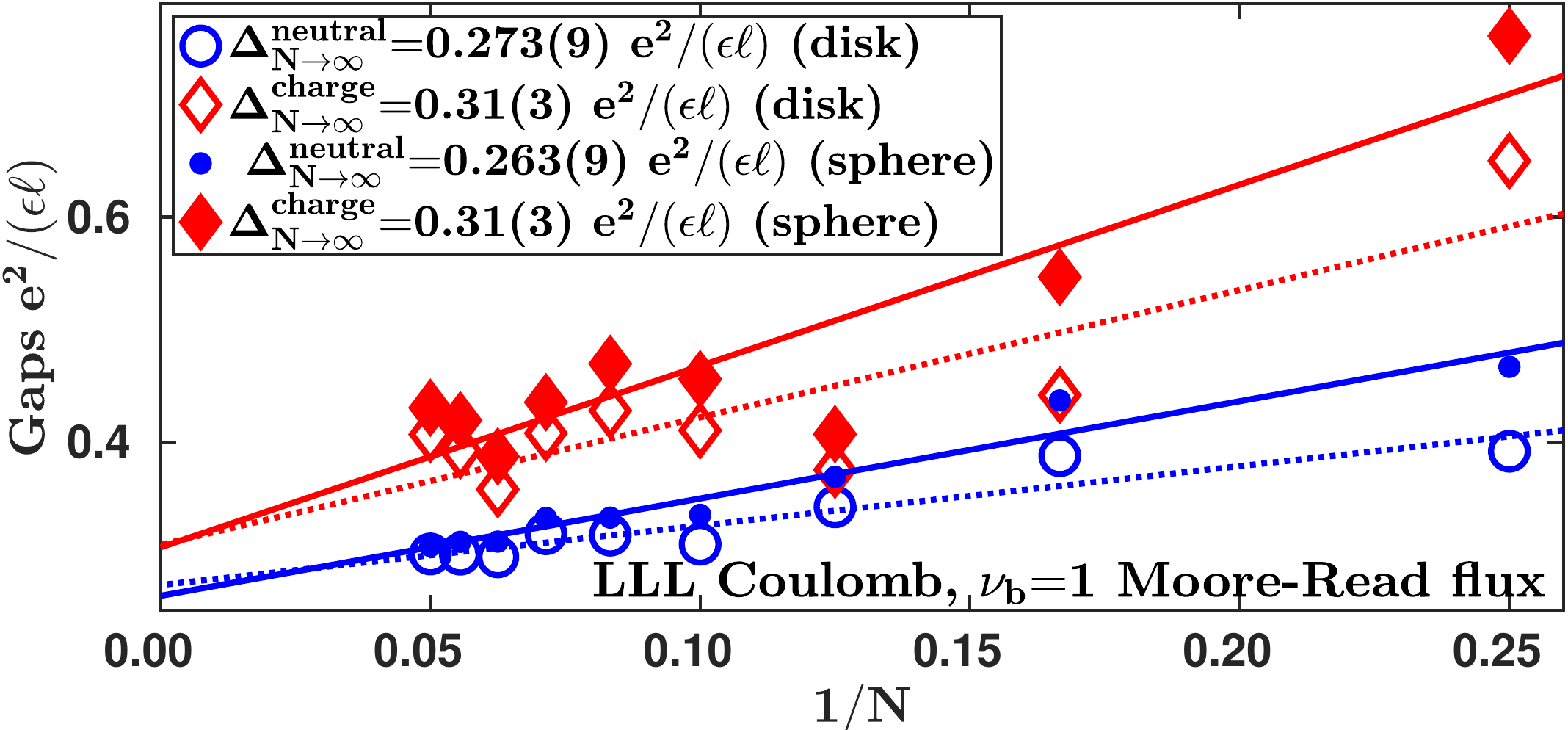} \\
		\caption{(color online) Thermodynamic extrapolation of the neutral (blue circles) and charge (red diamonds) gaps of the $V_{0}$ (top panel) and LLL Coulomb interactions (bottom panel) obtained in the spherical geometry using the disk (open symbols) and spherical (filled symbols) pseudopotentials for $N$ electrons at the bosonic $\nu_{b}{=}1$ MR Pf flux of $2Q{=}N{-}2$. The lines show a linear extrapolation of the gaps as a function of $1/N$ and the extrapolated gaps are quoted in the plots with the error in the extrapolation shown in the parentheses. These include gaps previously given in Refs.~\cite{Regnault03}.} 
		\label{fig: nu_b_1_bosonic_MR_Pf_gaps_V0_LLL_Coulomb}
	\end{center}
\end{figure}

In Fig.~\ref{fig: nu_b_1_bosonic_MR_Pf_gaps_V0_LLL_Coulomb} we show the neutral and charge gaps for the $V_{0}$ and Coulomb interactions evaluated this way at $\nu_{b}{=}1$. We find that the gaps for both interactions are sizable and of the same order since the Coulomb interaction is dominated by $V_{0}$ and $V^{\rm disk,~Coulomb}_{0}{=}0.886$. Furthermore, the Coulomb gaps obtained from the disk and spherical pseudopotentials are fully consistent. These results suggest that the $\nu_{b}{=}1$ bosonic MR-phase can be stabilized by the hard-core $V_{0}$ and Coulomb interactions in the LLL. Some of the gaps shown in Fig.~\ref{fig: nu_b_1_bosonic_MR_Pf_gaps_V0_LLL_Coulomb}, for systems smaller than the largest ones considered here, were previously given in Refs.~\cite{Regnault03}. 

\section{Periodic Interaction}
\label{appx-periodic-int}
The Coulomb interaction in real space with finite width corrections can be written as
\begin{equation}
V_{\rm C}(r)=\int dw_1 \int dw_2\frac{|\xi(w_1)|^2| \xi(w_2)|^2}{\sqrt{r^2+(w_1 - w_2)^2}}
\end{equation}
where $r^2=(x_1-x_2)^2+(y_1-y_2)^2$  and $r$ is the in-plane distance. $\xi(w)$ represents the wave function of electrons in the transverse direction and $w$ is the transverse coordinate. However, on a torus, the interactions are periodic i.e.
\beq
V(r+mL_1+nL_2) = V(r)
\eeq
where $m$ and $n$ are integers. We use the periodic form of the interaction given by
\begin{equation}
\label{per-int}
V(r) = \frac{1}{L^2\rm{Im} (\tau)}\sum_q \tilde{V}_{\rm C}(q)e^{i\vec{q}\cdot \vec{r}}
\end{equation}
\beq
\vec{q} = \left( {2\pi m \over L}, -{2\pi \tau _1 m \over L \tau _2} +{2\pi  n \over L \tau _2} \right)
\eeq
 where $\tilde{V}_{\rm C}(q)$ is obtained by taking Fourier transform of the interaction $V_{\rm C}(r)$.
\begin{widetext}
\begin{eqnarray}
\tilde{V}_{\rm C}(q)&=&\int d^2r ~e^{-i\vec{q}\cdot \vec{r}} \left( \int dw_1 \int dw_2\frac{|\xi(w_1)|^2| \xi(w_2)|^2}{\sqrt{r^2+(w_1 - w_2)^2}} \right) \nonumber \\
&=& \int dw_1 \int dw_2~ |\xi(w_1)|^2| \xi(w_2)|^2 \int d^2r ~e^{-i\vec{q}\cdot \vec{r}}\frac{1}{\sqrt{r^2+(w_1 - w_2)^2}} \nonumber \\
&=& \int dw_1 \int dw_2 ~|\xi(w_1)|^2| \xi(w_2)|^2 \int_0 ^{\infty} dr~r \frac{1}{\sqrt{r^2+(w_1 - w_2)^2}}2\pi J_0(qr) \nonumber \\
&=&  \int dw_1 \int dw_2 ~|\xi(w_1)|^2| \xi(w_2)|^2 \left(\frac{2\pi}{q}\right)e^{-q|w_1-w_2|}
\end{eqnarray}
\end{widetext}
For our calculations, we use a cutoff value of $|m|,|n|\leq 30$ in Eq.~\eqref{per-int}. We neglect the $q=0$ term in Eq.~\eqref{per-int} since it cancels the electron-background and background-background energies.  We also need to include the self-interaction energy, which is the interaction of an electron in the principal zone with its image in other zone. The form of the self-interaction energy for the Coulomb interaction in the LLL is given by \cite{Yoshioka83,Bonsall77} :
\beq
W & = & -{e^2\over \epsilon \sqrt{L^2\abs{\tau}}} \left[ 2 - \sum _{mn}' \varphi_{-{1 \over 2}}(\pi(|\tau| m^2 + |\tau| ^{-1}n^2))\right] \nonumber \\
\varphi _{n}&=&\int _1^{\infty} dt e^{-zt}t^n 
\eeq
The primed summation excludes the term $m=0,n=0$. At a given system size with similar periodic boundary conditions, the same self-interaction  energy is independent of the state for a given QW width and density. 
\end{appendix}


%

\end{document}